\documentclass[11pt]{article}

\usepackage[utf8]{inputenc}
\usepackage[T1]{fontenc}
\usepackage[english]{babel}
\usepackage{lmodern}
\usepackage{xcolor}

\usepackage[colorlinks=False]{hyperref} 

\usepackage{amsmath}  
\usepackage{amsfonts,dsfont}
\usepackage{amssymb}
\usepackage{amsthm}
\usepackage{bm} 
\usepackage{comment}

\usepackage{float}
\usepackage[utf8]{inputenc}
\usepackage[T1]{fontenc}
\usepackage[english]{babel}
\usepackage{lmodern}
\usepackage{graphicx}
\usepackage{rotating}

\newlength{\bredde}
\def\slash#1{\settowidth{\bredde}{$#1$}\ifmmode\,\raisebox{.15ex}{/}
\hspace*{-\bredde} #1\else$\,\raisebox{.15ex}{/}\hspace*{-\bredde} #1$\fi}
\begin{document}

\title{A tridiagonal matrix-valued process with stochastic resetting for arbitrary Dyson index $\beta>0$}
\author{{Gernot Akemann$^1$, Satya N. Majumdar$^2$ and Patricia P\"a{\ss}ler$^1$}\\~\\
$^1$Faculty of Physics,
Bielefeld University,
Postfach 100131,
D-33501 Bielefeld, Germany\\
$^2$
LPTMS, CNRS, Univ. Paris-Sud, Universit\'e Paris-Saclay, 91405 Orsay, France
}
\maketitle
\begin{abstract}

We introduce a symmetric tridiagonal matrix-valued process ($\beta$-TMP) $H(t)$ whose diagonal entries 
$H_{k,k}(t)$ evolve independently via an Ornstein-Uhlenbeck process starting at the origin and the off-diagonal 
entries $H_{k,k+1}(t)$ evolve independently via the Cox-Ingersoll-Ross (CIR) process, starting at the origin and 
with parameters that depend on the row index $k$. We show that the joint distribution of the entries of the 
matrix can be computed exactly at all times and moreover, the joint distribution of its $N$ real eigenvalues can 
also be computed exactly at all times. We then subject this time-evolving matrix-valued process to stochastic 
resetting with rate $r$ in two different settings: (i) when the matrix entries are simultaneously reset to the 
origin with rate $r$ ($\beta$-SRTMP process) and (ii) when the matrix entries are independently reset to the 
origin with rate $r$ ($\beta$-IRTMP process). We show that the joint distribution of the eigenvalues of the 
$\beta$-SRTMP process at long times can be computed analytically and it coincides with the joint distribution of 
the positions of the resetting Dyson Brownian motion in its stationary state for arbitrary $\beta>0$. For the 
$\beta$-IRTMP stationary ensemble, computing analytically the joint distribution of eigenvalues or even the 
average density of eigenvalues is difficult. However, generating the stationary $\beta$-IRTMP ensemble 
numerically is relatively straightforward and we compare its numerical average eigenvalue density to the 
corresponding analytical results for the $\beta$-SRTMP stationary ensemble with same parameter values, showing 
that they are quite different from each other. Finally, we provide a simple and concrete application of this 
tridiagonal matrix-valued process in computing the annealed partition function of a disordered quantum 
tight-binding Hamiltonian on a one-dimensional lattice.

\end{abstract}

\maketitle

\section{Introduction}\label{intro}

Consider $N$ point particles on a line whose coordinates 
${\vec x} (t)\equiv \{x_1(t),x_2(t),\ldots, x_N(t)\}$ evolve stochastically
with time $t$ via the overdamped Langevin equation
\begin{equation}
\frac{dx_i}{dt}= - \mu\,x_i + D\, \sum_{j(\ne i)} \frac{1}{x_i-x_j} + \sqrt{\frac{2D}{\beta}}\, \eta_i(t),
\label{dbm0.def}
\end{equation}
where the $\eta_i(t)$ are independent zero-mean Gaussian white noises with correlator
\begin{equation}
\langle \eta_i(t)\eta_j(t')\rangle = \delta_{i,j}\delta(t-t')\, .
\label{noise.0}
\end{equation}
Here $\delta_{i,j}$ is the Kronecker delta function, namely $\delta_{i,i}=1$ for all $i=1,2\ldots, N$, while
$\delta_{i,j}=0$ for $j\ne i$.
In Eq. (\ref{dbm0.def}), the quantities $\{\mu, D, \beta\}$ are just three positive parameters. The
first term on the right hand side (rhs) represents a force due to a confining harmonic potential
$V(x_i)= \mu x_i^2/2$ acting on each particle that tries to drive the particle towards the origin. 
The second term on the rhs corresponds to a pairwise repulsion
term between the particles that tries to push them apart. The third term on the rhs corresponds to
thermal fluctuations induced by a heat bath with inverse temperature $\beta$ to which the particles are connected.
The stochastic process in Eq. (\ref{dbm0.def}), introduced by Dyson~\cite{Dyson1962a,Dyson1962b}, 
is referred to as the $\beta$-Dyson Brownian motion ($\beta$-DBM), 
with the inverse temperature $\beta$ usually referred to as the Dyson index. 
For simplicity, we assume that the Langevin equation \eqref{dbm0.def} starts from
the initial condition 
where all the particles are localised close to the origin, e.g.,
over an interval $(-\epsilon,\epsilon]$ with
\begin{equation}
x_i(0)= \epsilon\, \left(-1 +\frac{2i}{N}\right)\, , \quad i=1,2,\ldots N\, ,
\label{init.0}
\end{equation}
and eventually the limit $\epsilon\to 0$ is taken.
However, irrespective of the initial condition,
it is easy to see that at long times $t\to \infty$,
the joint probability density function (JPDF) $P_0(\vec x,t)$ of the positions of the particles approaches
a stationary form which is simply given by the equilibrium Gibbs-Boltzmann distribution
\begin{equation}
P_0^{\rm st}(\vec x)= P_0\left(\vec x, t\to \infty\right)\propto e^{-\beta\, E(\vec x)}\, ,
\label{gibbs.0}
\end{equation}
where the energy $E(\vec x)$ of the gas of particles is given by
\begin{equation}
E(\vec x)= \frac{\mu}{2D}\, \sum_{i=1}^N x_i^2 -\frac{1}{2}\sum_{i\ne j} \ln |x_i-x_j|\, .
\label{energy.0}
\end{equation}
The first term on the rhs of \eqref{energy.0} represents the external potential energy felt by the particles, while
the second term represents the pairwise repulsive interaction between the particles. Such a
gas of one-dimensional particles in thermal equilibrium is called the Dyson log-gas~\cite{F10}. One can rewrite the stationary
distribution \eqref{gibbs.0} in a more familiar form
\begin{equation}
P_0^{\rm st}(\vec x)=\frac{1}{Z_N(\beta)}\, \left(\frac{\beta \mu}{D}\right)^{N/2 +\beta N(N-1)/4}\, 
e^{-\frac{\beta \mu}{2D} \sum_{i=1}^N x_i^2}\, \prod_{i>j} |x_i-x_j|^{\beta}\, ,
\label{stat.0}
\end{equation}
where $Z_N(\beta)$ is a normalizing constant such that $\int_{\vec x}\, 
d\vec x\,P_0^{\rm st}(\vec x)=1$~\cite{F10}, i.e.,
\begin{equation}
Z_N(\beta)= \int_{-\infty}^{\infty}\cdots \int_{-\infty}^{\infty} dy_1\, dy_2\, \cdots dy_N\, e^{- \frac{1}{2}\, \sum_{i=1}^N y_i^2}\, \prod_{i>j} 
|y_i-y_j|^{\beta}=\frac{\Gamma\left(1+\frac{\beta}{2}\right)^N}{(2\pi)^{N/2}
\prod_{k=1}^{N}\Gamma\left( 1+\frac{\beta k}{2}\right)}\, .
\label{pf.0}
\end{equation} 
In this form \eqref{stat.0}, one recognises that for $\beta=1,2$ and $4$, this is just the joint distribution
of the $N$ real eigenvalues of an $(N\times N)$ Gaussian matrix that is either real symmetric $(\beta=1)$,
complex Hermitian $(\beta=2)$ or quaternionic self-dual $(\beta=4)$~\cite{Dyson1962a,Dyson1962b,F10}.
For general $\beta>0$, the result \eqref{stat.0} represents the JPDF of the $N$ real eigenvalues
of the tridiagonal Dumitriu-Edelman (DE) ensemble~\cite{DE2002}.

At any finite time $t$, for the three special values $\beta=1,\,2$ and $4$, Dyson showed~\cite{Dyson1962a,Dyson1962b}
that the positions $\{x_i(t)\}$ in \eqref{dbm0.def}
can be interpreted as $N$ real eigenvalues of a time-dependent $(N\times N)$ Gaussian matrix $H(t)$, 
which is real symmetric ($\beta=1$), complex Hermitian $(\beta=2)$ and quaternionic self-dual ($\beta=4$).
The entries of this matrix are evolving stochastically in a `fictitious' time $t$ by performing
independent Ornstein-Uhlenbeck (OU) processes. Such a time-dependent stochastic matrix $H(t)$ will
be referred to as a {\em matrix-valued process}. Dyson demonstrated, considering the evolution
of $H(t)$ from $t$ to $t+\Delta t$ and using a second order perturbation theory in $\Delta t$, that for
$\beta=1,2$, and $4$, if one diagonalizes this matrix $H(t)$ at any instant $t$, then
its $N$ real instantaneous eigenvalues $\vec{x}(t)$ evolve
via Eq. (\ref{dbm0.def}). However, the stochastic process \eqref{dbm0.def} is well defined
for any $\beta>0$. This $\beta$-DBM process is interesting in its own right (irrespective
of its connection to random matrix theory) and has been studied extensively in many different 
applications~\cite{F10,KT04,Bor09,ABG12,ABMV13,BBMP2014,AD2015,LY2017,GMS21,MM21,TLS23,BMS2025,FLO2025}. 
A natural question
then arises whether, for arbitrary $\beta>0$ (not necessarily $\beta=1,2$ and $4$), 
one can find an appropriate underlying matrix-valued process $H(t)$ 
with stochastically evolving entries such that its eigenvalues at any time $t$ evolve by the
$\beta$-DBM process \eqref{dbm0.def}.

This question becomes particularly relevant in the context of stochastic resetting
in many-body classical systems. Stochastic resetting simply means interrupting the natural dynamics (stochastic
or deterministic) of a system at random times and restarting the dynamics from the same initial condition~\cite{EM2011,EM2011_b}.
For the reader's benefit, we provide a brief introduction to stochastic resetting later in Section \ref{SR}.
One of the principal effects of stochastic resetting, amongst others, is to drive the system at long times
to a nonequilibrium stationary state (NESS), since the stochastic interruption typically violates detailed balance~\cite{EM2011}.
Recently, the $\beta$-DBM process in Eq. (\ref{dbm0.def}) in the presence of a simultaneous stochastic resetting
with rate $r$ was studied in Ref.~\cite{BMS2025}. More precisely, one starts the $\beta$-DBM process from the initial
condition \eqref{init.0} and evolves it up to a random time $\tau_1$ drawn from an exponential distribution
$p(\tau)= r\, e^{-r \tau}$. At this time $\tau_1$, all the particles are simultaneously and instantaneously 
reset to their initial positions in \eqref{init.0} and they again evolve up to a random time $\tau_2$ drawn
again from $p(\tau)$. Then they are reset and the process continues. 
This process was referred to as $\beta$-RDBM (resetting Dyson Brownian motion).
It was proved in Ref.~\cite{BMS2025} that
the JPDF of the positions of particles $P_r(\vec x, t)$, at long times $t\to \infty$, approaches a NESS 
$P_r^{\rm st}(\vec x)$ given explicitly, for arbitrary $\beta>0$, by
\begin{equation}
P_r^{\rm st}(\vec x)= \frac{r}{Z_N(\beta)}\,\int_0^{\infty} \frac{d\tau\, e^{-r\, \tau}}{
\left[\sigma(\tau)\right]^{N+ \beta N (N-1)/2}}\,
e^{-\frac{1}{2\, \sigma^2(\tau) }\sum_i x_i^2} \, \prod_{i> j}^N
|x_i-x_j|^{\beta}\,  ,
\label{RDBM_st.0}
\end{equation}
where 
\begin{equation}
\sigma^2(\tau)= \frac{D}{\beta\, \mu}\, \left(1- e^{-2\, \mu\, \tau}\right)\, 
\label{sigma2_def.0}
\end{equation}
and $Z_N(\beta)$ is given in Eq. (\ref{pf.0}). 
In the limit $r\to 0$, the JPDF $P_r^{\rm st}(\vec x)$ in Eq. (\ref{RDBM_st.0}) reduces to
$P_0^{\rm st}(\vec x)$ in Eq. (\ref{stat.0}). This can be easily seen by  implementing the change of variables $r\tau \to y$ in Eq. \eqref{RDBM_st.0}.

For $\beta=1, 2$ and $4$, since at any instant $t$ Eq. (\ref{dbm0.def})
describes the evolution of the eigenvalues of a Gaussian matrix with independent OU entries, it follows that the $\beta$-RDBM process describes the evolution of the eigenvalues of a resetting matrix-valued process whose
OU entries undergo simultaneous resetting to the origin with rate $r$~\cite{BMS2025}. Consequently, for these three
special values of $\beta=1, 2$ and $4$, this resetting matrix-valued process
also reaches a NESS as $t\to \infty$, i.e., a nonequilibrium ensemble (no longer Gaussian, but still 
rotationally invariant) whose
eigenvalues are jointly distributed via the JPDF $P_r^{\rm st}(\vec x)$ in Eq. (\ref{RDBM_st.0}). It is then
natural to ask if, for arbitrary $\beta>0$ (not necessarily $\beta=1,2$ and $4$), one can construct a 
resetting matrix-valued process $H(t)$ such that the matrix ensemble, driven by simultaneous resetting of its entries, 
approaches a NESS and its eigenvalues 
are jointly distributed via Eq. (\ref{RDBM_st.0}) with general $\beta>0$. The goal of this paper is to precisely construct
such a general-$\beta$ resetting matrix-valued process $H(t)$, thus generalising the results of Ref.~\cite{BMS2025}.

Let us briefly outline our main results. We first construct a tridiagonal matrix-valued process in fictitious time $t$
where the diagonal elements evolve independently as OU processes, while the off-diagonal elements perform independent
Cox-Ingersoll-Ross (CIR) processes~\cite{CIR85} (sometimes also known
as a special case of Feller process~\cite{Feller51}) with row-dependent parameters, with all entries starting from the origin. We call this
tridiagonal matrix-valued process the $\beta$-TMP process (tridiagonal matrix process). We then subject this
time evolution to stochastic resetting with rate $r$ whereby all matrix entries are simultaneously and instantaneously
reset to the origin with rate $r$. We call this resetting matrix-valued process as $\beta$-SRTMP (simultaneously
resetting tridiagonal matrix process). We show that at long times,
this $\beta$-SRTMP matrix process approaches a NESS, i.e., a stationary ensemble whose eigenvalues are precisely distributed
via Eq. (\ref{RDBM_st.0}) for arbitrary $\beta>0$. We also briefly introduce another resetting matrix ensemble,
where the independent entries of the tridiagonal $\beta$-TMP process undergo {\it independent} resetting, as opposed to
the simultaneous resetting in $\beta$-SRTMP process. We refer to this ensemble as $\beta$-IRTMP (independently
resetting tridiagonal matrix process). While the $\beta$-IRTMP process also approaches a NESS (different from that
of $\beta$-SRTMP) at long times, we were not able to find the JPDF of its eigenvalues in the NESS. It turns
out that numerically, it is much easier to generate the NESS ensemble of $\beta$-IRTMP since the matrix elements
remain independent at all times (similar to Wigner matrices in random matrix theory~\cite{F10}), compared to
the $\beta$-SRTMP process. Hence, we are able to compute numerically the average density of eigenvalues 
in the NESS ensemble of $\beta$-IRTMP and we demonstrate that they are quite different from the average
density of eigenvalues in the NESS of $\beta$-SRTMP that are analytically computable from the exact JPDF
in Eq. (\ref{RDBM_st.0}). Finally, we also provide a concrete application of the $\beta$-SRTMP 
NESS ensemble in the context of a single quantum particle with nearest neighbour disordered hopping rates on a one-dimensional
ring as in the celebrated Anderson model. We show that the annealed free energy in this random hopping model
can be computed exactly knowing the stationary JPDF in Eq. (\ref{RDBM_st.0}).    

The rest of the paper is organized as follows. In Section \ref{SR}, we provide a brief description of stochastic resetting
for readers who are unfamiliar with it. 
We then briefly recall in Section \ref{RDBM} some properties of the $\beta$-DBM process in Eq. (\ref{dbm0.def}) and then of
the $\beta$-RDBM process once the simultaneous resetting with rate $r$ is switched on.
In Section \ref{CIR}, we introduce the CIR process of a single particle and
derive its exact time-dependent position distribution at all times $t$. This result is crucially needed in
the next Section \ref{TMP}, where we first
introduce the matrix-valued $\beta$-TMP process for arbitrary $\beta>0$.
In Section \ref{RTMP}, we introduce the resetting matrix-valued processes
$\beta$-SRTMP (simultaneous resetting of entries) and $\beta$-IRTMP (independent resetting
of entries). We
prove that the JPDF of the eigenvalues of the $\beta$-SRTMP ensemble at long times is given precisely by
Eq. (\ref{RDBM_st.0}) for arbitrary $\beta>0$. We also present 
numerical results for the
average density of eigenvalues in the $\beta$-IRTMP stationary ensemble and compare it to the analytical results
for the corresponding $\beta$-SRTMP stationary ensemble, demonstrating that they are quite different from each other.
Section \ref{Hopping} contains an application of our results in the disordered Anderson-type hopping model
on a one-dimensional lattice. Finally, we conclude with a summary and outlook
in Section \ref{Conclusion}.

\section{Stochastic resetting: a brief introduction}
\label{SR}

Since stochastic resetting plays a central role in this paper, we briefly outline here the main ideas behind it for
the benefit of the readers.
The subject of stochastic resetting, introduced in Ref.~\cite{EM2011}, has seen an explosion of activities
both theoretically and experimentally, spanning across disciplines such as physics, chemistry, biology,
mathematics and computer science (for reviews see ~\cite{EMS2020,GJ22,PKR22}). The main idea 
behind stochastic resetting
is very simple. Imagine a process evolving under its own natural dynamics (deterministic or stochastic) starting
from a fixed initial condition. 
Stochastic resetting simply means interrupting the natural dynamics at random times and re-starting the process
from the same initial condition. The original motivation for studying this process came from random search problems~\cite{EM2011,EM2011_b}.
Imagine a searcher searching for a fixed target in space by its own random movement, e.g., a diffusive searcher.
The mean time to find the target can be very long depending on the dynamics, because there are `rogue' trajectories
that can take the searcher away from the target. It turns out to be beneficial if instead the searcher returns
to its starting point with a certain probability or resetting rate $r$ (in case of continuous-time dynamics)
and restarts the process. The rational behind this is simple: the resetting move cuts off `rogue' wandering
trajectories and also lets the walker explore new regions by starting a new trajectory. Moreover, for a diffusive
searcher, it was found that there is, in fact, an optimal resetting rate $r^*$ that minimizes the mean search time~\cite{EM2011,EM2011_b}.
This aspect of stochastic resetting has proved to be very useful in designing random search algorithms
with a stochastic resetting component and led to several important practical applications~\cite{EMS2020,PKR22}.

There is also a second aspect of stochastic resetting on which we will mainly focus here. 
Stochastic resetting typically drives a system
to a nonequilibrium stationary state (NESS) at long times by eliminating long wandering trajectories. 
This steady state is `nonequilibrium' because the resetting move violates detailed balance: there is
always a probability current from any position to the initial position via instantaneous resetting move,
but there is no reverse move. Hence, there is always a nonzero probability current in configuration
space even in the steady state~\cite{EM2011,EMS2020}. 

The physical properties in the resetting driven NESS have since been studied in both single particle
and multiple particles (with or without interactions) systems, with a host of interesting phenomena (for reviews see \cite{EMS2020,PKR22,GJ22,NG23}). 
The many-body reset induced NESS has been studied theoretically
in a number of systems that include fluctuating $(1+1)$-dimensional interfaces~\cite{GMS2014}, symmetric exclusion
process in one dimension~\cite{BKP19}, $d$-dimensional Ising model evolving via the 
Glauber dynamics~\cite{MMS2020,CZ2024,AN2026},
spatio-temporally chaotic systems under resetting~\cite{AK2025},
Brownian and non-Brownian
particles with independent resettings~\cite{SSM2021,VAM2022,DHMMRS2023,VM2026,MV2026}
and also for the Dyson 
Brownian motion~\cite{BMS2025}. More recently, it was found that even non-interacting
systems, when driven to a NESS by simultaneous resetting, may exhibit 
nontrivial dynamically 
emergent correlations (DEC) (see Ref. \cite{MS2026} for
a recent review on DEC). Correlations between the particles in these systems emerge from the
shared history of resetting events. Several models of DEC have been recently studied 
which include
independent Brownian~\cite{BLMS2023,BMS2023} and non-Brownian processes such as L\'evy flights~\cite{BLMS2024} 
undergoing simultaneous Poissonian resetting with
constant rate $r$ and also simultaneous non-Poissonian resetting~\cite{MBMS25} or simultaneous 
partial resetting~\cite{Galla26}, 
independent Brownian particles
with simultaneous resetting induced by first-passage events~\cite{BMP2025,BMS2026} etc. 
Recently, an information
theoretic interpretation has been discussed to shed light on the mathematical nature of these DEC~\cite{Olsen2026}. 
Simultaneous resetting introduces nontrivial correlations between particles even
in systems that do not reach a NESS at late times, e.g., for independent non-Markovian `monkey walks' with preferential relocation
to previously visited trajectories~\cite{BM2026}. 
Going beyond the simultaneous resetting of all degrees of freedom, 
the effect of resetting has also been studied in systems where only a subset of the degrees of freedom reset,
such as in subsystem resetting in many-body systems such as in the Kuramoto model~\cite{MCG2024,AMG2025} or 
in the recently introduced `batch resetting' model for
independent Brownian particles~\cite{MMS2026}. The dynamically emergent correlations 
in the NESS between independent Brownian particles in a harmonic trap whose stiffness changes
between two values $\mu_1$ and $\mu_2$ with some rates have also been studied both
theoretically~\cite{BKMS2024} and experimentally in colloidal systems in a laser trap~\cite{BCKMPS2025}.
In this model, the limit $\mu_1\to \infty$ corresponds to resetting all the particles simultaneously
to the origin of the harmonic trap~\cite{BKMS2024}.

For simultaneous resetting in a generic many-body system,
there exists a very simple general relation, based on a renewal argument, that
connects the probability distribution in the configuration space in the presence and in the absence of resetting.
This argument is very general and holds for any process. 
Consider a system with any number of degrees of freedom
and let us denote a configuration of the system by $C$. The system undergoes its natural dynamics in continuous time
starting from a fixed configuration $C_0$
and let $r$ denote the resetting rate at which the system returns instantaneously from a current configuration
to its initial configuration $C_0$. The resetting events thus form a Poisson process with rate $r$.
Let $P_r(C,t)$ denote the probability that the system is in configuration $C$ at time $t$, starting
at $C_0$ at time $0$. The subscript $r$ denotes the presence of a nonzero resetting rate $r$. 
This is actually the propagator $P_r(C,t)= G_r(C,t|C_0,0)$ from $C_0$ at $t=0$ to $C$ at time $t$, but we omit
the explicit dependence on $C_0$ for brevity. Similarly, $P_0(C,t)$  
denotes the probability that the reset-free system ($r=0$) is in configuration $C$ at time $t$, starting from $C_0$
at time $0$. These two
probabilities are simply related to each other by the following renewal relation~\cite{GMS2014,EMS2020}
\begin{equation}
P_r(C,t)= e^{-r t}\, P_0(C,t) + r\, \int_0^t d\tau\, e^{-r\, \tau}\, P_0(C,\tau)\, .
\label{renewal_gen.1}
\end{equation}
The interpretation of the two terms on the rhs of Eq. (\ref{renewal_gen.1}) is very simple.
Consider all possible trajectories of the system going from $C_0$ at time $0$ to the configuration $C$ at time $t$. 
There are two possibilities: (i) no resetting events in $[0,t]$ and (ii) one or more 
resetting events occur in $[0,t]$. The probability for event (i) is simply $e^{-r t}$ since the resetting events 
are Poissonian with rate $r$.
In this case, the probability to reach $C$ at time $t$ (starting at $C_0$) is simply the bare (reset-free) probability
$P_0(C, t)$ (also starting at $C_0$).
Hence, multiplying $P_0(C,t)$ by $e^{-r t}$ gives the joint probability
of reaching $C$ at time $t$ and that there is no resetting event in $[0,t]$. This explains the first term
on the rhs of Eq. (\ref{renewal_gen.1}).
Now, consider the event (ii) when there are one or more
resettings in $[0,t]$. In this case, it is useful to first consider the epoch $t_l$ at which the last resetting before
$t$ occurred in the trajectory. Let $\tau=t-t_l$. This is useful because conditioned on the last resetting at $t_l$
before $t$, the process evolves freely (without resetting) during the interval $\tau$ from $t_l$ to $t$.
The probability for the process to reach $C$ at time $t$ (starting from $C_0$), conditioned on $t_l$, is then
$P_0(C, \tau=t-t_l)$ where we assumed that the propagator of the reset-free process depends only on the 
time difference $\tau=t-t_l$.
Now the probability of having the last resetting event at $t_l$ is simply the
probability of a reset at $t_l$ (which occurs with density $r$) followed by no resetting during
$\tau=t-t_l$ that occurs with probability $e^{-r \tau}$. Hence, multiplying all factors and integrating $\tau$
over $[0,t]$ gives the joint probability of reaching $C$ at time $t$ and having one or more resettings in $[0,t]$.
This explains the second term in Eq. (\ref{renewal_gen.1}). It is easy to check 
that $P_r(C, t)$ is normalized to unity
when summed over all configurations $C$. Thus, the renewal equation \eqref{renewal_gen.1} 
allows one to compute the probability $P_r(C, t)$ of the process
exactly at any time $t$, provided one knows the probability $P_0(C, t)$ of the underlying
reset-free process {\em at all times}. 

\vskip 0.3cm

\noindent{ \bf Nonequilibrium stationary state (NESS).} In the limit $t\to \infty$, the first term on the rhs of Eq. (\ref{renewal_gen.1})
drops out and we get the stationary state
\begin{equation}
P_r^{\rm st}(C)= P_r(C, t\to \infty)= r \int_0^{\infty} d\tau\, e^{-r \tau}\, P_0(C, \tau)\, ,
\label{stat_gen.1}
\end{equation}
provided, of course, the integral in Eq. (\ref{stat_gen.1}) exists. Thus the NESS is fully determined if one
knows the probability $P_0(C,t)$ of the underlying reset-free process {\it at all times}, and not just at late times. In many situations, it is easy to determine the late time behaviour of the reset-free propagator
$P_0(C,t)$, but is much harder to compute $P_0(C,t)$ {\em at all times} $t$ which is needed to determine the stationary
probability $P_r^{\rm st}(C)$ of the process with reset. 

As a simple example, consider the underlying reset-free process to represent the position of a one-dimensional
Brownian motion with position $x(t)$ at time $t$, starting from the origin at time $t=0$~\cite{EM2011}. 
In this example, the configuration $C$ is characterized by the
position $x$ of the particle which is the only degree of freedom. Since the process starts from the origin,
the probability density of the reset-free Brownian motion is simply
\begin{equation}
P_0(x,t)= \frac{1}{\sqrt {4\,\pi\, D\, t}}\, e^{-x^2/4Dt}\, ,
\label{1d_Brown.1}
\end{equation}
where $D>0$ represents the diffusion constant. If now this walker is subjected to resetting to the origin with a constant rate $r$,
the position distribution $P_r(x,t)$ at long times approaches a NESS. Substituting (\ref{1d_Brown.1}) in Eq. (\ref{stat_gen.1}) one
obtains the exact stationary probability density with resetting~\cite{EM2011}
\begin{equation}
P_r^{\rm st}(x)= r\, \int_0^{\infty} d\tau\, e^{-r\, \tau}\, \frac{1}{\sqrt{4\, \pi\, D\, \tau}}\, e^{-x^2/4D\tau}=
\frac{1}{2}\, \sqrt{\frac{r}{D}}\, e^{-\sqrt{\frac{r}{D}}\, |x|} \, .
\label{1d_rbm.1}
 \end{equation}
Thus the stationary position distribution is non-Gaussian, and decays exponentially over a length scale $\xi=\sqrt{D/r}$
that represents the typical distance the walker explores between two consecutive resetting events.

Another example consists of studying a set of $N$ independent Brownian motions, all starting 
at the origin and resetting
simultaneously to the origin with rate $r$~\cite{BLMS2023}. This is also equivalent to a single Brownian motion in $N$
dimensions that resets to the origin with a constant rate $r$~\cite{EM2014,BLMS2023}. In this case, the configuration 
$C\equiv \{x_1,x_2,\ldots, x_N\}\equiv {\vec x}$, where $x_i$ represents the position of the $i$-th particle in one dimension.
Since the particles are independent, the joint probability distribution of the positions of $N$ particles without reset
at any time $t$ is simply given by the product
\begin{equation}
P_0(\vec x, t)= \prod_{i=1}^N\frac{1}{\sqrt{4\, \pi\, D\, t}}\, e^{-x_i^2/{4Dt}}\, .
\label{Nd_Brown.1}
\end{equation}
Substituting (\ref{Nd_Brown.1}) in Eq. (\ref{stat_gen.1}), one obtains the stationary position distribution of
the process with reset~\cite{EM2014,BLMS2023}
\begin{eqnarray}
P_r^{\rm st}(\vec x)&= &r\, \int_0^{\infty} d\tau\, e^{-r\, \tau}\, \prod_{i=1}^N \frac{1}{\sqrt{4\,\pi\, D\, \tau}}\,
e^{-x_i^2/{4D\tau}} \nonumber \\
&= &  \left(\frac{r}{2\pi D}\right)^{N/2}\, \left(\sqrt{\frac{r}{D}}\, R\right)^{1-N/2}\, K_{N/2-1}\left( \sqrt{\frac{r}{D}}\, R\right)\, ,
 {\rm with}\quad R= \sqrt{x_1^2+x_2^2\ldots+ x_N^2}\, ,
\label{Nd_rbm.1}
\end{eqnarray}
where $K_\nu(z)$ is the modified Bessel function~\cite{GR_book}. For $N=1$, one recovers the result
in Eq. (\ref{1d_rbm.1}), as 
\begin{equation}
K_{-1/2}(z)=\sqrt{\frac{\pi}{2z}}\,\exp[-z], \quad {\rm with}\quad z\ge 0\, .
\label{K-1/2}
\end{equation}
Thus, for any $N\ge 1$, to determine fully
the stationary state in the presence of resetting, we need the full time-dependence of the reset-free propagator.
In the next section, we will apply this general theory of resetting to the $\beta$-DBM process defined in 
Eq. (\ref{dbm0.def}).

\section{Dyson Brownian motion and underlying matrix-valued process}
\label{RDBM}

In this section, we briefly recall the $\beta$-DBM process \eqref{dbm0.def} and its underlying matrix-valued process,
first without resetting and then in the presence of a simultaneous resetting with rate $r>0$.

\subsection{$\beta$-DBM process and the underlying matrix-valued process}

Consider first the standard $\beta$-Dyson Brownian motion ($\beta$-DBM) defined by
the Langevin equation \eqref{dbm0.def}. As discussed in the introduction, for three special values of the
parameter $\beta$ ($\beta=1,2$ and $4$),   
Dyson showed~\cite{Dyson1962a,Dyson1962b} that the set of positions $\{x_i(t)\}$ can be interpreted 
as the $N$ real eigenvalues
of a time-dependent $(N\times N)$ Gaussian matrix $H(t)$ (real symmetric, complex Hermitian and 
quaternion self-dual for $\beta=1,2,4$ respectively), whose elements
are evolving via the Ornstein-Uhlenbeck (OU) process. 
For instance, for $\beta=1$, the real symmetric matrix entry $H_{j,k}(t)$ evolves
via
\begin{equation}
\frac{d H_{j,k}(t)}{dt}= - \mu\, H_{j, k}(t) + \sqrt{\left(1+ \delta_{j,k}\right)\, D}\, \xi_{j, k}(t)\, ,
\label{OU.1}
\end{equation}
where the $\xi_{j,k}(t)$ are again independent (for $j\geq k$ subject to symmetry) zero-mean Gaussian white noises with correlator
$\langle \xi_{j,k}(t)\xi_{j', k'}(t')\rangle= \delta_{j, j'}\, \delta_{k,k'}\, \delta(t-t')$.
A similar construction can be made for $\beta=2$ and $\beta=4$ with complex 
and quaternionic matrix entries, subject to the Hermiticity and self-duality condition, respectively.  
The JPDF of the matrix entries, starting from $H_{j,k}(0)=0$ (for simplicity), 
then evolve as a Gaussian process (up to an overall normalization constant)
\begin{equation}
P_0(H,t)= {\rm Prob.}\left[H(t)=H\right]\propto \frac{1}{\left[\sigma(t)\right]^{N+\beta\, N(N-1)/2}}\, 
\exp\left[- \frac{1}{2\, \sigma^2(t)}\, {\rm Tr}(H^2) \right]\ , 
\label{matrix_OU.1}
\end{equation}
where we recall
\begin{equation}
\sigma^2(t)= \frac{D}{\beta\, \mu}\, \left(1- e^{-2\,\mu\, t}\right)\, .
\label{sigma2_def}
\end{equation}
Notice that when writing out the invariant probability distribution of matrix elements \eqref{matrix_OU.1}, it factorises into the product of 
$N+\beta\, N(N-1)/2$ independent real Gaussian factors, with different variances for the diagonal 
and off-diagonal matrix elements, due to symmetry. For example, for $\beta=1$, we have for the 
$N(N+1)/2$ independent real matrix elements $H_{l\geq k}$
\begin{equation}
{\rm Prob.}\left[H(t)=H\right]_{\beta=1}\propto \frac{1}{\left[\sigma(t)\right]^{N(N+1)/2}}\, 
\exp\left[- \frac{1}{2\, \sigma^2(t)}\, \sum_{j=1}^N\, H_{j,j}^2 - \frac{1}{\sigma^2(t)}\, \sum_{j>k}^N
H_{j,k}^2 \right]\ .
\label{matrix_OU.1b1}
\end{equation}

At late times, the JPDF \eqref{matrix_OU.1} of the matrix entries then converges to the standard, 
stationary (st) Gaussian ensembles for $\beta=1,\, 2,\, 4$ given by
\begin{equation}
P_0^{\rm st}[H]= \lim_{t\to \infty}{\rm Prob.}\left[H(t)=H\right]\propto \exp\left[- 
\frac{\beta\, \mu}{2D}\, {\rm Tr}(H^2)\right]\, .
\label{GE_st}
\end{equation}
At any instant of time $t$, if one diagonalises the time-dependent Gaussian matrix $H(t)$ with OU entries and
with $\beta=1,2,4$, the instantaneous eigenvalues (all real) $\vec{x}(t)$ evolve
via Eq. (\ref{dbm0.def}), as was shown by Dyson~\cite{Dyson1962a,Dyson1962b}. 
One can then ask: what is the JPDF of the eigenvalues $P_0(\{x_i\},t)\equiv P_0(\vec x,t)$ 
at time $t$ for $\beta=1,2,4$? This can be answered using a simple rescaling approach, without doing any explicit
computation, as shown below.

Consider first the long-time limit.
In this limit, since the matrix process approaches a Gaussian ensemble, 
the JPDF of the eigenvalues also approaches the standard form in Eq. (\ref{stat.0}). 
The subscript $0$ in $P_0$ in Eq. (\ref{stat.0})
just indicates that the process is reset-free. In fact, the JPDF of the eigenvalues at any instant $t$
can also be deduced immediately, for $\beta=1,2,4$, by using a simple rescaling argument. 
Rescaling the matrix $H(t)$ by $\sigma(t)$, i.e., 
defining a matrix $\tilde{H}(t)= H(t)/\sigma(t)$ with $\sigma(t)$ defined in Eq. (\ref{sigma2_def}), one sees from
Eq. (\ref{matrix_OU.1}) that
at any instant $t$, the JPDF of the matrix $\tilde{H}(t)$ has a time-independent form
\begin{equation}
{\rm Prob.}\left[\tilde{H}(t)=\tilde{H}\right]\propto 
\exp\left[- \frac{1}{2}\, {\rm Tr}({\tilde{H}}^2)\right]\, .
\label{matrix_OU.2}
\end{equation}
Consequently, from Eqs. (\ref{GE_st}) and (\ref{stat.0}), it follows that 
the eigenvalues $\tilde{\vec{x}}$ of $\tilde{X}$ are
also time-independent with a JPDF given by
\begin{equation}
P_0\left(\tilde{\vec{x}}\right)=
\frac{1}{Z_N(\beta)}\, e^{-\frac{1}{2}\sum_{i=1}^N \tilde{x}_i^2} \, \prod_{ i>j}^N
|\tilde{x}_i-\tilde{x}_j|^{\beta} \, ,
\label{st_jpdf.2}
\end{equation}
Reverting back to the unscaled eigenvalues $x_i(t)= \sigma(t)\, \tilde{x}_i$, one sees immediately that the PDF
of the original unscaled eigenvalues $\vec{x}(t)$ at any time $t$ is given by~\cite{F10}
\begin{equation}
P_0\left(\vec x, t\right)=
\frac{1}{Z_N(\beta)}\, \frac{1}{[\sigma(t)]^{N+ \beta N(N-1)/2}}\, 
e^{-\frac{1}{2 \sigma^2(t) }\sum_{i=1}^N x_i^2} \, \prod_{i> j}^N
|x_i-x_j|^{\beta} \, , \quad {\rm for}\quad \beta=1,\, 2,\, 4\, ,
\label{dbm_jpdf.2}
\end{equation}
where we recall that $\sigma(t)$ is given in Eq. (\ref{sigma2_def}). Note that the overall factor 
$[\sigma(t)]^{-N-\beta N(N-1)/2}$ comes from the change of variables in the Vandermonde determinant 
to the power $\beta$, but crucially $Z_N(\beta)$ does not depend
on $\sigma(t)$. In deriving the result in Eq. (\ref{dbm_jpdf.2}), we assumed that 
$\beta=1,\, 2$ or $4$ so that we could exploit the underlying matrix-valued process
and its rotational symmetries for $\beta=1,\,2,\, 4$. 
We will see below that the result in \eqref{dbm_jpdf.2} actually holds 
for arbitrary $\beta>0$~\cite{BMS2025}, and not necessarily for $\beta=1,\,2,\, 4$.

\vskip 0.3cm

\noindent{\bf Propagator for the $\beta$-DBM process at arbitrary time and arbitrary $\beta>0$.}
Consider now the $\beta$-DBM process in Eq. (\ref{dbm0.def}) for arbitrary $\beta>0$. From the discussion
above, we have seen that
for $\beta=1,2,4$, the interacting particle positions $\vec{x}(t)$ evolving via Eq. (\ref{dbm0.def}) can
be interpreted as the eigenvalues of an underlying  matrix OU process $H(t)$ and the JPDF of $\vec{x}(t)$ at any time $t$
can be computed explicitly as in Eq. (\ref{dbm_jpdf.2}). This is simply done
by exploiting the connection to the underlying matrix OU process and a simple rescaling argument. 
One natural question is: can one compute the JPDF of $\vec{x}(t)$ evolving 
via \eqref{dbm0.def} (i.e., for the $\beta$-DBM process) at
all times for a generic $\beta \ne 1,\, 2,\, 4$? By solving explicitly the Fokker-Planck
equation associated to the Langevin equation \eqref{dbm0.def}, it was shown recently~\cite{BMS2025} that 
actually, the time-dependent JPDF in Eq. (\ref{dbm_jpdf.2}) is exact for any $t$ and any 
$\beta>0$, not necessarily 
restricted to $\beta=1,2,4$. In this derivation, it was assumed that the 
Langevin process \eqref{dbm0.def} starts
from the initial condition \eqref{init.0} with $\epsilon\to 0$.
Hence the JPDF \eqref{dbm_jpdf.2}, for arbitrary $\beta>0$, can also be interpreted as the
propagator of the $\beta$-DBM process from $\vec x (0)\equiv \{x_1(0)\to0, x_2(0)\to 0,\ldots, x_N(0)\to 0\}$ to 
$\vec x \equiv \{x_1,x_2,\ldots, x_N\}$ in time $t$ 
\begin{equation}
P_0\left(\vec x, t\right)=\frac{1}{Z_N(\beta)}\, 
\frac{1}{[\sigma(t)]^{N+ \beta N(N-1)/2}}\, e^{-\frac{1}{2 \sigma^2(t) }\sum_i x_i^2} \, \prod_{i> j}^N
|x_i-x_j|^{\beta} \, , 
\label{prop.1}
\end{equation}
with $\sigma(t)$ from Eq. \eqref{sigma2_def} and $Z_N(\beta)$ is given in Eq. (\ref{pf.0}).

\subsection{Resetting Dyson Brownian Motion ($\beta$-RDBM) and its underlying matrix-valued process}

So, why are we interested in computing the exact JPDF of $\vec{x}(t)$ of the $\beta$-DBM process at all times? 
As discussed in Section \ref{SR}, this knowledge at all times is indeed needed to determine the exact NESS of this
process when it is subjected to simultaneous resetting of all the particles to the initial condition with rate $r$.
To be more precise, consider the $\beta$-DBM process in Eq. (\ref{dbm0.def})
and imagine resetting the process back to its initial condition \eqref{init.0} with a constant rate $r$. 
The initial condition $\vec x=\vec 0$ (in the limiting sense $\epsilon\to 0$ in \eqref{init.0}) 
corresponds to resetting all the particles simultaneously to the origin with rate $r$.
We recall that in the presence of a finite resetting rate $r$, this process has been named 
resetting Dyson Brownian 
motion ($\beta$-RDBM)~\cite{BMS2025}.
Let us ask: what is the JPDF $P_r(\vec x,t)$ of the positions of the particles at time $t$ of 
this $\beta$-RDBM process? 
The subscript $r$
denotes the resetting rate $r$. For $r=0$ (in the absence of resetting), the JPDF $P_0(\vec x, t)$ 
is given exactly by Eq. (\ref{prop.1}).
The JPDF $P_r(\vec x,t)$ in the presence of resetting can then be obtained by using the general renewal 
equation (\ref{renewal_gen.1})
and it reads~\cite{BMS2025}
\begin{equation}
P_r(\vec x, t) = e^{-r t}\, P_0(\vec x, t) + r\, \int_0^t d\tau\, e^{-r\, \tau}\, P_0(\vec x, \tau)\, ,
\label{renewal.1}
\end{equation}
where $P_0(\vec x, t)$ is the propagator of the reset-free process given explicitly in Eq. (\ref{prop.1}). 

In particular, at late times $t\to \infty$, the JPDF of the $\beta$-RDBM process in Eq. (\ref{renewal.1})
approaches a stationary form (the first term in Eq. (\ref{renewal.1}) drops out 
as $t\to \infty$) given by~\cite{BMS2025}
\begin{equation}
P_r^{\rm st}(\vec x)= P_r(\vec x, t\to \infty)=r\, \int_0^{\infty} d\tau\, e^{-r\, \tau}\, P_0(\vec x, \tau)\, ,
\label{RDBM_st.1}
\end{equation}
with $P_0(\vec x, \tau)$ given in Eq. (\ref{prop.1}). Substituting the explicit form of $P_0(\vec x, \tau)$ 
from Eq. (\ref{prop.1}),
one then gets the stationary JPDF of $\beta$-RDBM process in Eq. (\ref{RDBM_st.0})
in the reset induced NESS for arbitrary $\beta>0$~\cite{BMS2025}.

\vskip 0.3cm

\noindent {\bf Summary so far and a natural question.} 
So, what did we learn so far? One of the interesting conclusions from the above exercise is the following.
Consider the three special values $\beta=1,2,4$ for which there is an underlying OU matrix.
If the matrix elements now undergo an OU process with resetting, i.e., starting at the origin, the matrix $H(t)$ 
evolves
by the OU process as in Eq. (\ref{OU.1}) up to a random time distributed exponentially with parameter $r$,
then it resets instantaneously to $H=0$ and then again evolves up to a random time etc.
So, the probability distribution of the matrix $H(t)$ in the presence of resetting with rate $r$ can 
be written again using the renewal structure in Eq. (\ref{renewal.1}) as
\begin{equation}
P_r[H(t)=H]= e^{-r\, t}\, P_0[H,t] + r \, \int_0^t d\tau\, e^{-r\, \tau}\, P_0[H, \tau]\, ,
\label{matrix_renewal.1}
\end{equation}
where $\tau=t-t_l$ is again the time difference between $t$ and the last resetting time $t_l$ before $t$. In
Eq. (\ref{matrix_renewal.1}), the reset-free propagator of the matrix $P_0[H,t]$ 
is given by Eq. (\ref{matrix_OU.1}).
In particular at late times, the JPDF of the resetting matrix-valued process in Eq. (\ref{matrix_renewal.1})
approaches a stationary distribution
\begin{equation}
P_r^{\rm st}(H) \propto r\, \int_0^{\infty} d\tau\, e^{-r\, \tau}\, \frac{1}{\left[\sigma(\tau)\right]^{N+
\beta\, N(N-1)/2}}\,
\exp\left[- \frac{1}{2\, \sigma^2(\tau)}\, {\rm Tr}(H^2) \right]\,,  
\label{matrix_reset_st.1}
\end{equation}
with $\sigma(t)$ from Eq. \eqref{sigma2_def}.
Here, the proportionality constant just normalizes the matrix PDF. Such a deformed Gaussian matrix ensemble is 
called super-statistical Gaussian ensemble and has been studied in the literature in other contexts~\cite{AM2005,BCP2008,AMAV2009}.
We then conclude that if a matrix $H$ is distributed via Eq. (\ref{matrix_reset_st.1}) with $\beta=1,2,4$, then
its eigenvalues $\{x_i\}$'s are distributed via Eq. (\ref{RDBM_st.0}). 
In other words the $\beta$-RDBM process, for $\beta=1,2,4$, has a stationary distribution \eqref{RDBM_st.0}
which can be interpreted as the JPDF of the eigenvalues drawn from a superstatistical matrix
ensemble \eqref{matrix_reset_st.1}.

A natural question then is: what happens when $\beta\neq 1,\, 2,\, 4$? In other words, 
for a generic $\beta>0$, is there an underlying 
matrix reset ensemble such that its eigenvalues are distributed according to the stationary JPDF
of the $\beta$-RDBM process \eqref{RDBM_st.0}? 
In Section \ref{TMP}, we will see that indeed, for arbitrary $\beta>0$, one can construct a matrix-valued resetting
process $H(t)$ that reaches a stationary ensemble at late times and the 
eigenvalues of that stationary ensemble are indeed distributed
via Eq. (\ref{RDBM_st.0}). 
But in order to construct explicitly such a matrix-valued process, we need to first remind the reader
about another stochastic process known as the Cox-Ingersoll-Ross (CIR) process, which we do
in the next section.

\section{Cox-Ingersoll-Ross (CIR) process: A reminder}
\label{CIR}

Consider a stochastic process $x(t)$ defined on the semi-infinite line $x\ge 0$ that evolves via the Langevin equation
\begin{equation}
\frac{dx}{dt}= \mu_0\, (\theta-x(t))+ \sigma_0\, \sqrt{x(t)}\, \eta(t)\, ,
\label{cir.1}
\end{equation}
starting from $x_0\ge 0$. Here $\eta(t)$ is a zero-mean Gaussian while noise with correlator
$\langle \eta(t)\eta(t')\rangle= \delta(t-t')$. The multiplicative term in Eq. (\ref{cir.1})
is interpreted in the It$\hat{\rm o}$ sense. The process is characterized by three parameters
$\mu_0>0$, $\theta> 0$ and $\sigma_0>0$.
This Langevin process in Eq. (\ref{cir.1}) is a particular case of the
celebrated Feller process~\cite{Feller51} that describes an overdamped particle
in the presence of a linear drift and a space dependent diffusion
constant. The process (\ref{cir.1})
has been studied extensively and has found numerous applications
across disciplines: from neurobiology~\cite{CR73,CR74},
finance~\cite{CIR85,Hull,Heston93,DY2002} and branching process with death~\cite{MR2025}.
In finance, the process in Eq. (\ref{cir.1}) goes by the name of CIR (Cox-Ingersoll-Ross)
process~\cite{CIR85}.
The first-passage properties and the extreme value statistics of the CIR process have been investigated more
recently~\cite{MP2012,Somrita2022,MR2025}.

Let $p_0(x,t|x_0)$ denote the probability density for the process to be at $x$ at time $t$, starting from $x_0>0$ at $t=0$.
The exact result for $p_0(x,t|x_0)$ is known~\cite{CR73,CIR85}, but we provide a derivation of this result from first principles, to be self-containing.
The probability density $p_0(x,t|x_0)$ (in short $p_0(x,t)$ for convenience) satisfies the Fokker-Planck equation
\begin{equation}
\frac{\partial p_0(x,t)}{\partial t}= \frac{\sigma_0^2}{2}\, \frac{\partial^2}{\partial x^2}(x\, p_0) - 
\frac{\partial }{\partial x}\left[ (\mu_0\, \theta-
\mu_0\, x)\, p_0\right]\, , \quad {\rm valid}\,\, {\rm for}\quad x\ge 0\, .
\label{FP_CIR.1}
\end{equation}
It satisfies the initial condition
\begin{equation}
p_0(x,t=0|x_0)= \delta(x-x_0)\, .
\label{IC_CIR.1}
\end{equation}
The boundary conditions are as follows. As $x\to \infty$, it follows from Eq. (\ref{FP_CIR.1}) that $p_0(x,t)\sim 
\exp[-2\mu_0\, x/\sigma^2]$. The boundary condition at $x=0$ is slightly tricky. We first note that Eq. (\ref{FP_CIR.1})
can be expressed as a continuity equation: $\partial_t p_0(x,t)= - \partial_x j(x,t)$, where the current density $j(x,t)$
is given by
\begin{equation}
j(x,t)= - \frac{\sigma_0^2}{2}\, \left[x\, \partial_x p_0(x,t) + 
\left( \left(1- \frac{2\mu_0\,\theta}{\sigma_0^2}\right)+ \frac{2\mu_0 x}{\sigma_0^2}
\right)\, p_0(x,t)\right]\, .
\label{curr.1}
\end{equation}
Now, the current must vanish at $x=0$, i.e., $j(x,t)=0$ as $x\to 0$ since we assume that there is no leakage
of probability through $x=0$. Setting $j(x,t)$ in Eq. (\ref{curr.1}) to zero as $x\to 0$,
it follows that
\begin{equation}
p_0(x,t) \sim x^{1-b}\, \quad {\rm as}\quad x\to 0\, , \quad \quad {\rm where} \quad b= 2- \frac{2\mu_0\, \theta}{\sigma_0^2} \, . 
\label{def_b}
\end{equation}
It turns out to be convenient to rescale the position $z= 2 \mu_0 x/\sigma_0^2$.  
Then $p_0(x,t)= \frac{2\mu_0}{\sigma_0^2}\, \tilde{p}_0(z,t)$ where $\tilde{p}_0(z,t)$, from Eq. (\ref{FP_CIR.1})  satisfies 
\begin{equation}
\partial_t \tilde{p}_0(z,t) = \mu_0\, \left[ z\, \partial_z^2 \tilde{p}_0(z,t) + (b+z)\, \partial_z 
\tilde{p}_0(z,t) +\tilde{p}_0(z,t)\right]
, \quad {\rm valid}\,\, {\rm for}\quad z\ge 0\, .
\label{FPz_CIR.1}
\end{equation}
where $b$ is given in Eq. (\ref{def_b}). From Eq. (\ref{IC_CIR.1}), it satisfies the initial condition
\begin{equation}
\tilde{p}_0(z,t=0|z_0)= \delta(z-z_0)\, \quad {\rm where}\quad z_0= \frac{2\mu_0 x_0}{\sigma_0^2}\, .
\label{ICz_CIR.1}
\end{equation}
In addition, the two boundary conditions at $x\to \infty$ and $x\to 0$ translate, in the rescaled coordinate $z$, to
\begin{eqnarray}
\tilde{p}_0(z\to \infty, t) & \sim & e^{-z} \label{bczl.1} \\
\tilde{p}_0(z\to 0, t) &\sim & z^{1-b} \, , \label{bcz0.1}
\end{eqnarray} 
with $b$ given in Eq. (\ref{def_b}).

The partial differential equation (\ref{FPz_CIR.1}) can be solved by the standard eigenfunction expansion method.
We first express the solution as the linear combination
\begin{equation}
\tilde{p}_0(z,t) = \sum_{\lambda} A_\lambda\, e^{-\lambda\, t}\, \psi_\lambda(z)\, ,
\label{eigen_exp.1}
\end{equation}
where $A_\lambda$'s are constants and
the eigenfunction $\psi_\lambda(z)$ satisfies the eigenvalue equation
\begin{equation}
z\, \frac{d^2 \psi_{\lambda}(z)}{dz^2} + (b+z)\, \frac{d\psi_\lambda(z)}{dz} + \left(1+ \frac{\lambda}{\mu_0}\right)\, \psi_\lambda(z)=0\, .
\label{ODE.1}
\end{equation}
in the region $z\ge 0$, with the boundary conditions (which follow from Eqs. (\ref{bczl.1}) and (\ref{bcz0.1}))
\begin{eqnarray}
\psi_\lambda(z\to \infty) &\sim & e^{-z} \label{bczl.2} \\
\psi_\lambda(z\to 0) & \sim & z^{1-b} \, . \label{bcz0.2}
\end{eqnarray}
To solve the eigenvalue equation (\ref{ODE.1}), we first make the substitution
\begin{equation}
\psi_\lambda(z)= e^{-z}\, \phi_\lambda(z)\, ,
\label{subst.1}
\end{equation}
in order to reduce it to a known differential equation. Indeed, with this substitution (\ref{subst.1}), the function
$\phi_\lambda(z)$ satisfies the standard Kummer differential equation~\cite{AS_book}
\begin{equation}
z\, \, \frac{d^2 \phi_{\lambda}(z)}{dz^2} + (b-z)\, \frac{d\phi_\lambda(z)}{dz}  -a\, \phi_\lambda(z)=0\, , \quad {\rm where}\quad
a= 1- \frac{2\mu_0\theta}{\sigma_0^2}- \frac{\lambda}{\mu_0}\, .
\label{ODE_phi}
\end{equation}
There are two linearly independent solutions to Eq. (\ref{ODE_phi}) denoted respectively by $U(a,b,z)$ and $M(a,b,z)$~\cite{AS_book}. 
Hence, the most general solution of Eq. (\ref{ODE.1}) can be written as
\begin{equation}
\psi_\lambda(z) = C_1\, e^{-z}\, U(a, b, z) +C_2\, e^{-z}\, M(a, b, z)\, , \quad {\rm where}
\quad a= 1- \frac{2\mu_0\theta}{\sigma_0^2}- \frac{\lambda}{\mu_0} \quad {\rm and}\quad b=2- \frac{2\mu_0\, \theta}{\sigma_0^2} \, .
\label{sol.1}
\end{equation}
Here $C_1$ and $C_2$ are arbitrary constants yet to be fixed. The solution (\ref{sol.1}) must satisfy the boundary conditions
(\ref{bczl.2}) and (\ref{bcz0.2}). 

Let us first consider the boundary condition as $z\to \infty$ in Eq. (\ref{bczl.2}). The function $M(a,b,z)$ and 
$U(a,b,z)$ have the following asymptotic behaviors as $z\to \infty$~\cite{AS_book}
\begin{eqnarray}
M(a,b,z) &\approx & \frac{\Gamma(b)}{\Gamma(a)}\, z^{a-b}\, e^z \label{mzl.1} \\
U(a,b,z) &\approx & z^{-a}\, . \label{uzl.1}
\end{eqnarray}
Using these asymptotic behaviors in the general solution (\ref{sol.1}), we see that for large $z$, the solution
will behave as $\sim C_2\, z^{a-b}$. However, the boundary condition 
in Eq. (\ref{bczl.2}) demands that the eigenfunction must decay as $\sim e^{-z}$ to leading order for large $z$. 
Hence, to satisfy this boundary condition, we must have
$C_2=0$. Hence the solution now reads
\begin{equation}
\psi_\lambda(z) = C_1\, e^{-z}\, U(a, b, z)\, ,
\label{sol.2}
\end{equation}
which then decays for large $z$ as $\sim z^{-a}\, e^{-z}$ as required. We now investigate the other boundary condition
(\ref{bcz0.2}) as $z\to 0$. Using the known asymptotic behaviour of $U(a,b,z)$ as $z\to 0$~\cite{AS_book}, one finds that
the solution in Eq. (\ref{sol.2}) behaves, for small $z$, as
\begin{equation}
\psi_\lambda(z)\approx C_1\, \left[ \frac{z^{1-b}}{\Gamma(a)}\, \left(a_0+a_1\, z+ a_2\, z^2+\cdots\right)
+ \frac{\Gamma(1-b)}{\Gamma(1+a-b)}\, \left(b_0+b_1\, z+ b_2\, z^2+\cdots\right)\right]\, ,
\label{psi_smallz.1}
\end{equation}
where $a_i$'s and $b_i$'s are known constants~\cite{AS_book}.
Thus, to satisfy the boundary condition (\ref{bcz0.2}), the second series in the small $z$ expansion 
in Eq. (\ref{psi_smallz.1}) must vanish identically. This is possible if and only if the denominator
$\Gamma(1+a-b)$ in the second series in Eq. (\ref{psi_smallz.1}) diverges, which happens provided
\begin{equation}
1+a-b= -n \quad {\rm where} \quad n=0,1,2,\ldots\, .
\label{quant.1}
\end{equation}
Substituting the exact expressions of $a$ and $b$ from Eq. (\ref{sol.1}), this condition translates
into the quantization condition
\begin{equation}
\lambda= \mu_0\, n\, , \quad {\rm with}\quad n=0,1,2,\ldots
\label{quant_cond}
\end{equation}
This then fully selects the discrete spectrum of the eigenvalues $\lambda$. Furthermore, with $\lambda=\mu_0\, n$,
the constant $a$ in Eq. (\ref{sol.1}) also gets quantized as
\begin{equation}
a= 1- \frac{2\mu_0\theta}{\sigma_0^2} -n \, .
\label{a_quant}
\end{equation}

Consequently, the eigenfunction associated with the eigenvalue $\lambda=\mu_0\, n$ is then given from Eq. (\ref{sol.2})
\begin{equation}
\psi_n(z)= c_n\, e^{-z}\, U\left( 1- \frac{2\mu_0 \theta}{\sigma_0^2}-n, 2- \frac{2\mu_0\theta}{\sigma_0^2}, z\right)\, ,
\label{sol.3}
\end{equation}
where $c_n$ is an overall constant yet to be determined. We next use the exact identity~\cite{GR_book}
\begin{equation}
U(b-1-n,b,z)\propto z^{1-b}\, L_n^{1-b}(z)\, ,
\label{iden.1}
\end{equation}
where $L_n^{\alpha}(z)$ is the generalised Laguerre polynomial. Consequently, the $n$-th eigenfunction can be
expressed as
\begin{equation}
\psi_n(z)= B_n\, z^{\gamma}\, e^{-z}\, L_n^{\gamma}(z)\, , \quad {\rm where}\quad 
\gamma= \frac{2\mu_0\theta}{\sigma_0^2}-1\, ,
\label{sol.4}
\end{equation}
and $B_n$ is an overall constant. Hence, using the eigenfunction expansion (\ref{eigen_exp.1}), the
full time-dependent solution can be expressed as
\begin{equation}
\tilde{p}_0(z,t) = \sum_{n=0}^{\infty} A_n\, e^{- \mu_0\,n\, t}\, z^{\gamma}\, e^{-z}\,
L_n^{\gamma} (z)\, , \quad {\rm where} \quad \gamma= \frac{2\mu_0\theta}{\sigma_0^2}-1\, ,
\label{full_sol.1}
\end{equation}
and $A_n$'s are arbitrary constants that get fixed from the initial condition $\tilde{p}_0(z,t=0)=\delta(z-z_0)$.
Indeed, setting $t=0$ in Eq. (\ref{full_sol.1}) gives
\begin{equation}
\delta(z-z_0)= \sum_{n=0}^{\infty} A_n\, z^{\gamma}\, e^{-z}\, L_n^{\gamma} (z)\, .
\label{init.2}
\end{equation}
We next make use of the orthonormality relations satisfied by the Laguerre polynomials~\cite{GR_book}, namely,
\begin{equation}
\int_0^{\infty}dz\, e^{-z}\, z^{\gamma}\, L_n^{\gamma}(z)\, L_m^{\gamma}(z) = \Gamma(1+\gamma)\, \binom{n+\gamma}{n}\, \delta_{n,m}\, .
\label{ortho_cond.1}
\end{equation}
Multiplying both sides of Eq. (\ref{init.2}) by $ L_m^{\gamma}(z)$ and integrating over $z$
upon using Eq. (\ref{ortho_cond.1}) fixes the constant $A_m$ as
\begin{equation}
A_m= \frac{\Gamma(m+1)}{\Gamma(m+\gamma+1)}\, L_m^{\gamma}(z_0)\, .
\label{Am_fixed}
\end{equation}
Substituting this expression of $A_m$ in Eq. (\ref{full_sol.1}) then provides the full explicit time-dependent
solution as
\begin{equation}
\tilde{p}_0(z,t) = \sum_{n=0}^{\infty} \frac{\Gamma(n+1)}{\Gamma(n+\gamma+1)}\, e^{-n\, \mu_0\, t}\, 
z^{\gamma}\, e^{-z}\, L_n^{\gamma}(z)\, L_n^{\gamma}(z_0)\, , \quad {\rm with}\quad 
\gamma= \frac{2\mu_0\theta}{\sigma_0^2}-1 \, .
\label{full_sol.2}
\end{equation}
This solution can be further simplified by using the known expression for the generating function of the
Laguerre polynomial~\cite{GR_book}
\begin{equation}
\sum_{n=0}^{\infty} \frac{\Gamma(n+1)}{\Gamma(n+\gamma+1)}\, L_n^{\gamma}(z)\, L_n^{\gamma}(z_0)\, \omega^n 
= \frac{(z\, z_0\, \omega)^{-\gamma/2}}{1-\omega}\, e^{- \frac{\omega}{1-\omega}\, (z+z_0)}\, I_\gamma\left( 
\frac{2}{1-\omega}\, \sqrt{z\, z_0\, \omega}\right)\, ,
\label{Laguerre_genf.1}
\end{equation}
where $I_\gamma(z)$ is the modified Bessel function of the first kind~\cite{GR_book}. Setting $\omega=e^{-\mu_0\, t}$
in Eq. (\ref{Laguerre_genf.1}) and using it in Eq. (\ref{full_sol.2}) gives a compact exact solution
\begin{equation}
\tilde{p}_0(z,t)= z^{\gamma}\, e^{-z}\, 
\frac{(z\, z_0\, \omega)^{-\gamma/2}}{1-\omega}\, 
e^{-\frac{\omega}{1-\omega}\, (z+z_0)}\, 
I_\gamma\left(\frac{2}{1-\omega}\, \sqrt{z\, z_0\, \omega}\right)\, , \quad {\rm where}\quad \omega= e^{-\mu_0\, t}\, .
\label{full_sol.3}
\end{equation}
It is easy to check, using Eq. (\ref{ortho_cond.1}), that $\tilde{p}_0(z,t)$ is normalized to unity (as it should be),
\begin{equation}
\int_0^{\infty}dz\, \tilde{p}_0(z,t)=1\, .
\label{check_norm}
\end{equation}

Finally, reverting back to the original coordinate 
$x= \sigma_0^2 z/(2\mu_0)$ and using $p_0(x,t)= 2\mu_0 \tilde{p}_0(z,t)/\sigma_0^2$, 
one finds the explicit time-dependent solution of the Fokker-Planck equation for the CIR process
\begin{equation}
p_0(x,t|x_0,0) \equiv p_0(x,t)= C(t)\, e^{-u-v} \left(\frac{v}{u}\right)^{\gamma/2}\, I_\gamma\left(2\, \sqrt{u\, v}\right)\, ,
\label{full_solx.1}
\end{equation}
where
\begin{equation}
C(t)= \frac{2\mu_0}{\sigma_0^2 \left(1- e^{-\mu_0\, t}\right)}\, , \quad\quad  u= C(t)\, x_0\, e^{-\mu_0\, t}\, ,
\quad\quad v= C(t)\, x\, , \quad {\rm and}\quad  \gamma= \frac{2\mu_0\theta}{\sigma_0^2}-1\, .
\label{def_cons}
\end{equation}

\vskip 0.3cm

\noindent{\bf Stationary solution.} In the limit $t\to \infty$, the probability density $p_0(x,t)$ approaches
a stationary solution, $p_0(x,t\to \infty) \to p^{\rm st}_0(x)$. To derive the stationary solution from Eq. (\ref{full_solx.1}),
we note from Eq. (\ref{def_cons}) that, to leading order as $t\to \infty$,
\begin{equation}
C(t)\to \frac{2\mu_0}{\sigma_0^2}\, , \quad\quad u\to \frac{2\mu_0 x_0}{\sigma_0^2}\, e^{-\mu_0\, t}\, ,
\quad\quad v\to \frac{2\mu_0\, x}{\sigma_0^2}\, .
\label{cons_st.1}
\end{equation}
Since $u\to 0$ in Eq. (\ref{full_solx.1}) when $t\to \infty$, we can use the small argument behavior of $I_\gamma(z)$~\cite{GR_book}
\begin{equation}
I_\gamma(z) \approx \frac{2^{-\gamma}}{\Gamma(\gamma+1)}\, z^{\gamma}\, \quad {\rm as} \quad z\to 0\,.
\label{BesselI_smallz}
\end{equation}
Substituting this behavior along with Eq. (\ref{cons_st.1}), one gets from Eq. (\ref{full_solx.1}) the limiting behavior as
$t\to \infty$, namely
\begin{equation}
p^{\rm st}_0(x)= p_0(x, t\to \infty)= \left(\frac{2\mu_0}{\sigma_0^2}\right)^{\gamma+1}\, \frac{x^{\gamma}}{\Gamma(\gamma+1)}\, 
e^{-2\mu_0\, x/\sigma_0^2}\, , \quad {\rm with}\quad \gamma=\frac{2\mu_0\theta}{\sigma_0^2}-1\, .
\label{stat_sol.1}
\end{equation}
It is easy to check that this stationary solution is normalized to unity, i.e., $\int_0^{\infty}dx\, p^{\rm st}_0(x)=1$, and
of course, it is independent of the initial position $x_0$.
Note that one can also derive this stationary solution directly from the eigenvalue expansion in Eq. (\ref{full_sol.2})
by retaining only the $n=0$ term that survives in the limit $t\to \infty$. The stationary solution $p^{\rm st}_0(x)$ 
in Eq. (\ref{stat_sol.1})
can be expressed as the probability density function (PDF) of a Gamma distributed random variable. 
We recall that a random variable $\Gamma(q,l)\ge 0$ is called Gamma distributed
if its PDF can be written as
\begin{equation}
F(x, q, l)= {\rm Prob.}\left[\Gamma(q,l)=x\right] = \frac{1}{\Gamma(q)\, l^q}\, x^{q-1}\, e^{-x/l}\, , 
\quad {\rm for}\quad x\ge 0\,  ,
\label{Gamma_pdf.1}
\end{equation}
where the shape factor $q\ge 0$ and the scale factor $l$ parametrize this PDF. Comparing Eq. (\ref{Gamma_pdf.1})
to the exact stationary solution $p_0^{\rm st}(x)$ in Eq. (\ref{stat_sol.1}), we see immediately that 
\begin{equation}
p^{\rm st}_0(x)= F\left(x,\, q=\frac{2\mu_0\theta}{\sigma_0^2},\, l= \frac{\sigma_0^2}{2\mu_0}\right)\, , 
\label{stat_sol.2}
\end{equation}
where $F(x,q,l)$ is given in Eq. (\ref{Gamma_pdf.1}). Thus 
$p^{\rm st}_0(x)={\rm Prob.}[y_{\rm st}=x]$ is the PDF of a Gamma distributed 
random variable $y_{\rm st}$
\begin{equation}
y_{\rm st}=\Gamma\left(q=\frac{2\mu_0\theta}{\sigma_0^2},\, l= \frac{\sigma_0^2}{2\mu_0}\right)\, .
\label{Gamma_random.1}
\end{equation}

\vskip 0.3cm

\noindent{\bf Full time-dependent solution for the special initial condition where $x_0=0$.}
For a generic initial condition $p_0(x, t=0)= \delta(x-x_0)$ with $x_0\ge 0$, the full time-dependent solution
(\ref{full_solx.1}) is somewhat complicated. However, for the special case when $x_0=0$, i.e., the process
starts at the origin, the full time-dependent solution takes a very simple self-similar form which will
be useful later. To derive this form, we note from Eq. (\ref{def_cons}) that as $x_0\to 0$, we have
$u\to 0$. Consequently, we can again use the small argument asymptotics of $I_\gamma(z)$ given in
Eq. (\ref{BesselI_smallz}) to take the limit $x_0\to 0$ in Eq. (\ref{full_solx.1}). This gives, for any $t\ge 0$,
\begin{equation}
p_0(x,t|x_0=0) = C(t)\, \frac{[C(t)\, x]^{\gamma}}{\Gamma(\gamma+1)}\, e^{-C(t)\, x}\, , \quad {\rm where}\quad
C(t)=\frac{2\mu_0}{\sigma_0^2 \left(1- e^{-\mu_0\, t}\right)}\, , \quad {\rm and}\quad
\gamma=\frac{2\mu_0\theta}{\sigma_0^2}-1\, .
\label{origin_init.1}
\end{equation}
This solution, for any $t\ge 0$ and $x\ge 0$, can be expressed in terms of the PDF $F(x,q,l)$ of a Gamma distributed
random variable, i.e.,
\begin{equation}
p_0(x,t|x_0=0)= F\left(x, \,  q=\frac{2\mu_0\, \theta}{\sigma_0^2},\,  l=\frac{1}{C(t)}=
\frac{\sigma_0^2 \left(1- e^{-\mu_0\, t}\right)}{2\mu_0}\right)\, ,
\label{sol_final.1}
\end{equation}
where $F(x,q,l)$ is defined in Eq. (\ref{Gamma_pdf.1}). Thus, $p_0(x,t|x_0=0)={\rm Prob.}[y(t)=x]$ represents the PDF of
a Gamma distributed random variable $y(t)$
\begin{equation}
y(t)= \Gamma\left(q=\frac{2\mu_0\, \theta}{\sigma_0^2},\,  l=
\frac{\sigma_0^2 \left(1- e^{-\mu_0\, t}\right)}{2\mu_0}\right)\, .
\label{yt_def.1}
\end{equation}
Of course, as $t\to \infty$, one recovers the stationary solution, i.e., $y(t\to \infty)\to y_{\rm st}$
with the latter given in Eq. (\ref{Gamma_random.1}).
In summary, for the special initial condition $x_0=0$, the PDF $p_0(x,t|x_0=0)$ of the CIR process is
given exactly, {\em at all times} $t\ge 0$, by the PDF of a Gamma distributed random variable defined in Eq. (\ref{yt_def.1}).

\section{A tridiagonal matrix-valued process ($\beta$-TMP)}
\label{TMP}

Having gathered the necessary background on the OU process and the CIR process, we now proceed to construct a matrix-valued
process \`a la Dyson. We consider a real, symmetric $(N\times N)$ tridiagonal matrix $H_{k,k'}(t)$ that evolves stochastically 
in a fictitious time $t$. Since the matrix is tridiagonal, by definition $H_{k,k'}(t)=0$ for all $t$, 
for $k' \ne k-1, k, k+1$. In addition, since the matrix is symmetric, we have $H_{k,k+1}(t)=H_{k+1,k}(t)$ for
$k=1,2,\ldots, N-1$. All nonzero matrix entries (not related by symmetry) are assumed to be
statistically independent of each other at all times $t$ and evolve by the following dynamics:

\begin{itemize}

\item{{\bf Diagonal elements $H_{k,k}(t)$.}} The $k$-th diagonal entry is assumed to evolve as an OU process, i.e., 
\begin{equation}
\frac{d H_{k,k}(t)}{dt}= -\mu\, H_{k,k}(t) + \sqrt{\frac{2D}{\beta}}\, \eta_k(t)\, ,
\label{TMP_diag.1}
\end{equation}
where $\eta_k(t)$'s are zero-mean Gaussian white noises with the correlator $\langle \eta_k(t)\eta_{k'}(t')\rangle= \delta_{k,k'}\, 
\delta(t-t')$. This process has three parameters $(\mu, \, D,\, \beta)$ which are arbitrary for the moment. We assume
that the process, for each $k=1,2\ldots, N$, starts at the origin, i.e., $H_{k,k}(t=0)=0$. The probability density of the process
$H_{k,k}(t)$ at time $t$ is given by the Gaussian distribution
\begin{equation}
{\rm Prob.}[H_{k,k}(t)=x]= \frac{1}{\sqrt{2\,\pi\,\sigma^2(t)}}\, e^{-\frac{x^2}{2\sigma^2(t)} }\, ,    
\label{OU_pdf.1}
\end{equation}
where
\begin{equation} \label{sigma_TMP}
	\sigma^2(t)=\frac{D}{\beta\, \mu}\, \left(1- e^{-2\,\mu\, t}\right)\, .
\end{equation}
Thus, $H_{k,k}(t)= {\cal N}\left(0, \sigma^2(t)\right)$ is just a zero-mean Gaussian normal variable 
with variance $\sigma^2(t)$.
For later convenience, let us express this after rescaling by a factor $1/\sqrt{2}$ as
\begin{equation}
H_{k,k}(t)=\frac{1}{\sqrt{2}}\, {\cal N}\left(0, 2\sigma^2(t)\right)\, ,
\label{Hdiag.1}
\end{equation}
with $\sigma(t)$ given in Eq. \eqref{sigma_TMP}.

\item{{\bf Off-diagonal elements $H_{k,k+1}(t)$.}} Consider the element $H_{k,k+1}(t)$ with $k=1,2,\ldots, N-1$.
We assume $H_{k,k+1}(t)= \sqrt{Y_k(t)/2}$ where $Y_k(t)$ represents a CIR process in Eq. (\ref{cir.1}) starting 
at the origin $Y_k(t=0)=0$
with parameters $(\mu_0,\, \theta_k,\, \sigma_0)$. From Eq. (\ref{yt_def.1}), $Y_k(t)$ is a Gamma distributed random variable
and hence we have\footnote{We choose a different labelling of the matrix elements $H_{j,k}$ here, compared to \cite{DE2002}.}
\begin{equation}
H_{k,k+1}(t)=\sqrt{\frac{\Gamma\left(q=\frac{2\mu_0\, \theta_k}{\sigma_0^2},\,  l=
\frac{\sigma_0^2 \left(1- e^{-\mu_0\, t}\right)}{2\mu_0}\right)}{2}}\, ,
\label{Hup.1}
\end{equation}
where we are at liberty to choose the parameters $(\mu_0, \theta_k, \sigma_0)$. We will make the following choice for the
parameters
\begin{equation}
\mu_0=2\, \mu\, , \quad\quad \theta_k= \frac{D}{\mu}(N-k)\, , \quad\quad {\rm and}\quad \sigma_0^2= \frac{8D}{\beta}\, ,
\label{para_choice}
\end{equation}
where $(\mu,\, D, \, \beta)$ are the parameters of the OU process characterizing the diagonal entries. 
With this choice of parameters, we then have
from Eq. (\ref{Hup.1})
\begin{equation}
H_{k,k+1}(t)= \sqrt{ \frac{\Gamma\left(\frac{\beta(N-k)}{2}\, ,\,  2\sigma^2(t)\right)}{2}}\, , 
\label{Hup.2}
\end{equation}
with $\sigma(t)$ given in Eq. \eqref{sigma_TMP}.
The probability density of the off-diagonal process $\sqrt{2}H_{k,k+1}(t)$ at time $t$ is given by the (generalized) Gamma distribution
\begin{equation}
	{\rm Prob.}\left[\sqrt{2}H_{k,k+1}(t)=x\right]= \frac{2^{1-\frac{\beta(N-k)}{2}}}{\Gamma\left(\frac{\beta(N-k)}{2}\right)\sigma^{\beta(N-k)}(t)}x^{\beta(N-k)-1}e^{-\frac{x^2}{2\sigma^2(t)}} , \text{ for } x>0.
\label{CIR_pdf.1}	
\end{equation}
\end{itemize}
This symmetric tridiagonal matrix $H(t)$ (with parameters $(\mu,\, D\, ,\beta)$) reads
\begin{equation}
\frac{1}{\sqrt{2}}	\begin{pmatrix}
		{\cal N}(0,2\sigma^2(t)) & \sqrt{\Gamma\left(\frac{\beta(N-1)}{2},2\sigma^2(t)\right)}& & &\\
		\sqrt{\Gamma\left(\frac{\beta(N-1)}{2},2\sigma^2(t)\right)} & {\cal N}(0,2\sigma^2(t))& \sqrt{\Gamma\left(\frac{\beta(N-2)}{2},2\sigma^2(t)\right)}& & \\
		\ddots & \ddots &\ddots&\\
			&  \sqrt{\Gamma\left(\frac{2\beta}{2},2\sigma^2(t)\right)} &
			{\cal N}(0,2\sigma^2(t))&\sqrt{\Gamma\left(\frac{\beta}{2},2\sigma^2(t)\right)}\\	
		& & \sqrt{\Gamma\left(\frac{\beta}{2},2\sigma^2(t)\right)}& {\cal N}(0,2\sigma^2(t))
	\end{pmatrix}.\!
	\label{H-matrix}
\end{equation}
The JPDF of the matrix elements of $H(t)$ is determined via Eqs. (\ref{OU_pdf.1}), (\ref{CIR_pdf.1})
\begin{equation}
P_0(H,t)=\frac{1}{c_N(\beta)}\frac{2^{-\beta N(N-1)/4}}{[
\sigma(t)]^{N+ \beta N(N-1)/2}} \prod_{j=1}^N\,
e^{-\frac{H_{j,j}^2}{2\sigma^2(t)}}\,
\prod_{k=1}^{N-1}\, H_{k,k+1}^{\beta(N-k)-1}\, e^{-\frac{H_{k,k+1}^2}{\sigma^2(t)}} \, , 
\label{JPDF_H}	
\end{equation}
where
\begin{equation}
	c_N(\beta)=
\frac{2^{N-1}}{(2\pi)^{N/2}\prod_{k=1}^{N-1}\Gamma\left(\frac{\beta k}{2}\right)}\, ,
\label{cnbeta_def}
\end{equation}
and $\sigma(t)$ is given in Eq. \eqref{sigma_TMP}.
The symmetric tridiagonal matrix $H(t)$ is a generalization of the tridiagonal matrix model 
introduced by Dumitriu-Edelman (DE)~\cite{DE2002}. The Chi-distribution of the off-diagonal 
model in the DE matrix ensemble is a special case of the Gamma-distribution when
the variance $\sigma^2(t)$ is time-independent and has unit value.
In the present case the variance $\sigma^2(t)$ in Eq. (\ref{cnbeta_def}), 
however, depends explicitly on time $t$.
The JPDF in terms of the real eigenvalues of $H(t)$ follows very 
similarly as in the time-independent model in~\cite{DE2002}. 
The exponential parts in Eq. (\ref{JPDF_H}) can be rewritten as
\begin{equation}
	\exp\left(-\sum_{k=1}^N\frac{H_{k,k}^2}{2\sigma^2(t)}-\sum_{k=1}^{N-1}\frac{H_{k,k+1}^2}{\sigma^2(t)}\right)=\exp \left(-\frac{1}{2\sigma^2(t)}{\rm Tr}(H^2)\right).
\end{equation}
It follows, that the JPDF of $H(t)$ is of the same functional form as in the time-independent model 
$H(0)$ (cf. Proof of Theorem 2.12 in ~\cite{DE2002}) up to the factor $1/\sigma^2(t)$ in the exponential 
and an overall normalization term incorporating $\sigma(t)$. To see this, 
we briefly review all ingredients for the proof. We impose a spectral decomposition of $H(t)$ as
\begin{equation}
	H(t)=QX(t)Q^t,
\end{equation}
where $X(t)={\rm diag}(x_1(t),x_2(t),\ldots,x_N(t))$ 
contains the real eigenvalues of $H(t)$ and $Q$ is a real orthogonal. We apply~\cite[Lemma 
2.5]{DE2002} and find that the first row $\vec{q}$ of $Q$ and the eigenvalues uniquely 
determine $Q$ and $H(t)$. Next, we use~\cite[Lemma 2.7]{DE2002} to find a rewriting of the 
Vandermonde determinant in terms of the off-diagonal elements $H_{k,k+1}(t)$ ($H_{k,k+1}(t)$ 
corresponds to $b_{N-k}$ in~\cite{DE2002}) and the spectral parameters $\vec{q}$. The 
computation of the Jacobian in~\cite[Lemma 2.9]{DE2002} for the change of variables from the 
matrix entries to the spectral parameters $(X(t),\vec{q})$ is unchanged, as the proof uses 
only the tridiagonal symmetric form of $H(t)$.

Hence, the joint distribution of the eigenvalues
$\vec{x}(t)$ at any instant $t$, and for any $\beta>0$, is given 
exactly by
\begin{equation}
P_0\left(\vec x, t\right)=
\frac{1}{Z_N(\beta)}\, 
\frac{1}{\left[\sigma(t)\right]^{N+ \beta\, N(N-1)/2}}\, 
e^{-\frac{1}{2\,\sigma^2(t)}\, \sum_{i=1}^N x_i^2} \, \prod_{i> j}^N
|x_i-x_j|^{\beta} \, ,
\label{tmp_jpdf.1}
\end{equation}
with $\sigma(t)$ from Eq. \eqref{sigma_TMP}.
The normalizing constant $Z_N(\beta)$ is given in Eq. (\ref{pf.0}) and most crucially, it
does not depend on $t$.

But this result (\ref{tmp_jpdf.1}) coincides precisely with the JPDF
of the $\beta$-DBM process in Eq. (\ref{prop.1}). Hence, this construction shows that for arbitrary $\beta>0$,
the $\beta$-DBM process corresponds to the eigenvalues of an underlying tridiagonal symmetric 
time-evolving matrix $H(t)$
whose diagonal entries perform independent OU processes starting at the origin and the off-diagonal entries
perform independent CIR processes (with a rank dependent shape parameter $q$ in Eq. (\ref{Hup.2})) starting at the origin.
We will call this tridiagonal matrix-valued process as $\beta$-TMP.

\section{Tridiagonal matrix-valued resetting processes}
\label{RTMP}

We introduced above the $\beta$-TMP process whose eigenvalues, at any time $t$, are distributed via the JPDF in 
Eq. (\ref{tmp_jpdf.1}) and coincides with the JPDF of $\beta$-DBM process in Eq. (\ref{dbm_jpdf.2}) for 
arbitrary $\beta>0$. We now subject the $\beta$-TMP process to Poissonian resetting with rate $r$. The resetting 
move can be introduced in two alternative ways: (i) $\beta$-SRTMP process where all the matrix entries {\em 
simultaneously} reset to the origin with rate $r$ and (ii) $\beta$-IRTMP process where the matrix entries reset 
{\em independently} with rate $r$ to the origin. Thus, in $\beta$-IRTMP process, the matrix entries (not 
connected by symmetry) remain independent at all times, while in $\beta$-SRTMP process the matrix entries get 
correlated with time due to the shared history of simutaneous resetting. We consider the two processes 
separately below.

\subsection{Tridiagonal matrix with simultaneous resetting of entries ( $\beta$-SRTMP)}

Our starting point is the exact time-dependent JPDF of the matrix entries $P_0(H(t))$ in
Eq. (\ref{JPDF_H}). If now the matrix entries are simultaneously reset to the origin with rate $r$,
the JPDF of the resetting matrix ensemble at time $t$ is given by the general renewal equation \eqref{renewal_gen.1}
 \begin{equation}
P_r(H,t)= e^{-r t}\, P_0(H, t) + r \int_0^t d\tau\, e^{-r \tau}\, P_0(H, \tau)\, ,
\label{SRTMP.1}
\end{equation}
where $P_0(H,t)$ is given in Eq. (\ref{JPDF_H}).
In particular, as $t\to \infty$, the $\beta$-SRTMP process converges to a stationary JPDF
\begin{equation}
P_r^{\rm st}(H)= r \int_0^{\infty} d\tau\, e^{-r \tau}\, P_0(H,\tau)\, .
\label{SRTMP_stat}
\end{equation}
Explicitly, it reads
\begin{equation}
P_r^{\rm st}(H)= \frac{r\, 2^{-\beta N(N-1)/4}}{c_N(\beta)}\, \int_0^{\infty} \frac{d\tau\, 
e^{-r\tau}}{\left[\sigma(\tau)\right]^{N+\beta N(N-1)/2}}\,\prod_{j=1}^N e^{-\frac{H_{j,j}^2}{2\sigma^2(\tau)}} \prod_{k=1}^{N-1} H_{k,k+1}^{\beta(N-k)-1}e^{-\frac{H_{k,k+1}^2}{\sigma^2(\tau)}}\, ,
\label{SRTMP_stat.2}
\end{equation}
where $c_N(\beta)$ and $\sigma^2(t)$ are defined in \eqref{cnbeta_def}. Given this distribution of the
matrix ensemble, one can then ask how are its eigenvalues distributed? When the matrix entries
are simultaneously reset with rate $r$, it follows that its eigenvlaues also undergo
simultaneous resetting with rate $r$. Now, at any given instant $t$, the JPDF
of the eigenvalues of the $\beta$-TMP process is given by Eq. (\ref{tmp_jpdf.1}).
We again apply the general renewal property \eqref{SRTMP_stat} in the limit $t\to \infty$, but now to the JPDF
of the eigenvalues rather than the matrix itself. This gives
the JPDF of the eigenvalues $\vec x\equiv 
\{x_1,x_2,\ldots, x_N\}$ of the $\beta$-SRTMP stationary ensemble in Eq. (\ref{SRTMP_stat.2}) as
\begin{equation}
P_r^{\rm st}(\vec x)= \frac{r}{Z_N(\beta)}\,\int_0^{\infty} \frac{d\tau\, e^{-r\, \tau}}{
\left[\sigma(\tau)\right]^{N+ \beta N (N-1)/2}}\,
e^{-\frac{1}{2\, \sigma^2(\tau) }\sum_i x_i^2} \, \prod_{i> j}^N
|x_i-x_j|^{\beta}\,
\label{SRTMP_eigen_st.0}
\end{equation}
where $Z_N(\beta)$ is given in Eq. (\ref{pf.0}) and we recall
\begin{equation}\label{sigma_RTMP}
	\sigma^2(\tau)=\frac{D}{\beta \mu}\,\left(1- e^{-2\, \mu\, \tau}\right)\, .
\end{equation}
Note that Eq. (\ref{SRTMP_eigen_st.0}) coincides with the JPDF of the $\beta$-RDBM process in Eq. (\ref{RDBM_st.0}). This
then answers one of our motivations raised in the introduction that there is indeed an underlying matrix-valued process, here
the $\beta$-SRTMP process, such that its eigenvalues are distributed according to \eqref{SRTMP_eigen_st.0} 
for arbitrary $\beta>0$.

The JPDF of the eigenvalues in Eq. (\ref{SRTMP_eigen_st.0}) for arbitrary $\beta>0$ was studied in detail
in Ref.~\cite{BMS2025} and several interesting exact results in the limit of large $N$ were derived. For instance, the average 
density of eigenvalues in the limit of large $N$ reads~\cite{BMS2025}
\begin{equation}
\label{eq:main-density}
\rho_N(x | r) \simeq \sqrt{\frac{\mu}{N D}} \;
f\left( x \sqrt{\frac{\mu}{N D}} , \, \frac{\mu}{r} \right) \;,
\end{equation}
where the scaling function $f(z, \nu= \mu/r)$ is symmetric around $z=0$ and
is independent of $\beta$. It has a
finite support over $z \in [-\sqrt{2}, \sqrt{2}]$ and is normalized to unity,
i.e., $\int_{-\sqrt{2}}^{\sqrt{2}}dz\, f(z, \nu) = 1$.
Its explicit expression reads~\cite{BMS2025}
\begin{equation} \label{eq:main-f}
f(z, \nu) = \frac{\Gamma\left( 1 + \frac{1}{2 \nu} \right)}{\sqrt{2 \pi}
\Gamma\left( \frac{3}{2}+\frac{1}{2 \nu}\right)}
\left( 1 - \frac{z^2}{2} \right)^{\frac{1 + \nu}{2 \nu}}\, \,
_2\,{F}_1 \left[1, \frac{1}{2 \nu}, \frac{3}{2} +
\frac{1}{2\nu}, 1 -\frac{z^2}{2} \right] , \quad z \in [-\sqrt{2},+\sqrt{2}] \;,
\end{equation}
where $\,_2\,{F}_1[a, b, c, z]$ is the hypergeometric function.
For some special values of $\nu=\mu/r$ the hypergeometric function simplifies,
for example for $\nu = 1$ and $\nu = 1/2$, where one obtains~\cite{BMS2025}
\begin{equation}
\label{eq:main-density-special}
f(z, \nu= 1) = \frac{1}{\sqrt{2}} - \frac{|z|}{2} \mbox{~~and~~}
f(z, \nu = 1/2) = \frac{2}{\pi} \left( \sqrt{2 - z^2} - |z| \,
{\rm ArcCos} \left[ \frac{|z|}{\sqrt{2}} \right] \right) \; ,
\end{equation}
see blue full lines in Figure \ref{fig-n20 nmtr50000} top respectively bottom row.
We end this section with a final remark. In the $\beta$-SRTMP process, the matrix elements are highly correlated
as seen in Eq. (\ref{SRTMP_stat.2}), as the $\tau$-dependent variance $\sigma^2(\tau)$, which is common to all independent matrix elements, is integrated over, creating correlations. Hence it is not easy to generate this matrix numerically. However, the
JPDF of the eigenvalues of this ensemble can be explicitly expressed as in Eq. (\ref{SRTMP_eigen_st.0}) which allows
for analytical computations.

\subsection{Tridiagonal matrix with independent resetting of entries ($\beta$-IRTMP)}

We now consider the $\beta$-IRTMP matrix process where each entry of the tridiagonal $\beta$-TMP process
resets independently to $0$ with rate $r$. Thus, in this case, the matrix elements remain independent at
all times, including in the stationary state. We now apply the general renewal equation \eqref{SRTMP_stat} in the limit $t\to \infty$
for the evolution of the each matrix entries separately. 
It follows that in the $t\to \infty$ limit, the elements
of the tridiagonal matrix $H(t)$ approach a stationary state, where the diagonal and the off-diagonal entries
are distributed as follows.

\begin{itemize}

\item{{\bf Diagonal elements.}} From Eqs. (\ref{OU_pdf.1}) and (\ref{stat_gen.1}), it follows
that the diagonal element $H_{k,k}(t)$, in the limit $t\to \infty$ and in the presence
of $r>0$, is distributed via the stationary PDF
\begin{equation}
\lim_{t\to \infty} {\rm Prob.}[H_{k,k}(t)=x]= r\, \int_0^{\infty} d\tau\, e^{-r\, \tau}\, \frac{1}{\sqrt{2\,\pi\,\sigma^2(\tau)}}\, 
e^{-\frac{x^2}{2\sigma^2(\tau)} }\, ,
\label{diag_pdf.1}
\end{equation}
with $\sigma(t)$ from Eq. \eqref{sigma_RTMP}.
The entries for different $k$ are independent of each other. 

\item{{\bf Off-diagonal elements.}} The matrix remains symmetric and tridiagonal. 
From Eqs. (\ref{Hup.1}) and (\ref{stat_gen.1}) it follows that 
the element $H_{k,k+1}(t)$, in
the limit $t\to \infty$ and $r>0$, can be expressed as
\begin{equation}
H_{k,k+1}(t\to \infty)= \sqrt{\frac{Y_k(r)}{2}}\, ,
\label{off_diag_element.1}
\end{equation}
where the random variable $Y_k(r)$ is Gamma distributed via the stationary PDF
\begin{equation}
{\rm Prob.}\left[Y_k(r)=y\right] = r\, \int_0^{\infty} d\tau\, e^{-r\, \tau}\, 
\frac{y^{q-1}}{\Gamma(q)\, \left[2\, \sigma^2(\tau)\right]^q}\, e^{-y/2\sigma^2(\tau)} ,
\label{off_diag_pdf.1}
\end{equation}
where 
\begin{equation}
q= \frac{\beta}{2}\,(N-k)\, ,
\label{off_diag_para.1}
\end{equation}
and $\sigma(t)$ given in Eq. \eqref{sigma_RTMP}.
\end{itemize}

The JPDF of matrix entries of this $\beta$-IRTMP in the stationary state is then given by the product of all 
independent entries of matrix $H(t)$ in the limit $t\to \infty$, 
\begin{equation}
{\rm Prob.}\left[ H, t\to \infty\right] \propto \lim_{t\to \infty}\, \left[ \prod_{j=1}^N {\rm Prob.}\left[ 
H_{j,j}(t)\right]\, 
\prod_{k=1}^{N-1} {\rm Prob.}\left[ H_{k,k+1}(t)\right] \right].
\label{prob.Htridiag}
\end{equation}
This can be compared for example with the distribution \eqref{matrix_OU.1b1} of independent elements for $\beta=1$
in the $t\to \infty$ limit. Notice, however, that here the JPDF in \eqref{prob.Htridiag} {\it cannot} be written 
in an invariant way, in terms of the trace of $H(t\to \infty)$ squared. Compared to the stationary JPDF with simultaneous resetting in \eqref{SRTMP_eigen_st.0}, where we integrate over the variable $\tau$ of the product of distributions of independent matrix elements with $\tau$-dependent variance $\sigma^2(\tau)$, we have in \eqref{prob.Htridiag} the product of integrals in the case of the independent resetting. In this sense, this $\beta$-IRTMP stationary
ensemble is like the Wigner matrices, rather than rotationally invariant matrices in random matrix theory~\cite{F10}.
In fact, given the stationary JPDF of the matrix entries of $\beta$-IRTMP ensemble in Eq. (\ref{prob.Htridiag}),
it is not easy to compute the JPDF of its eigenvalues, unlike in the $\beta$-SRTMP process in Eq. (\ref{SRTMP_eigen_st.0}).
In contrast, it is much easier to generate this matrix numerically since its entries are independent.
Below we compute numerically the average density of eigenvalues in the
$\beta$-IRTMP process and compare it with the exact analytical result in Eqs. (\ref{eq:main-density})
and (\ref{eq:main-f}) for the $\beta$-SRTMP process. We find they are quite different as expected.

\begin{figure}[h]
	\hspace{50pt}{$\beta=1$} \hspace{120pt}{$\beta=2.6$}\hspace{120pt}{$\beta=3.1$} \hfill\\
	\begin{turn}{90}\hspace{30pt}{$\mu=1$}\end{turn}
	\includegraphics[width=0.3\linewidth]{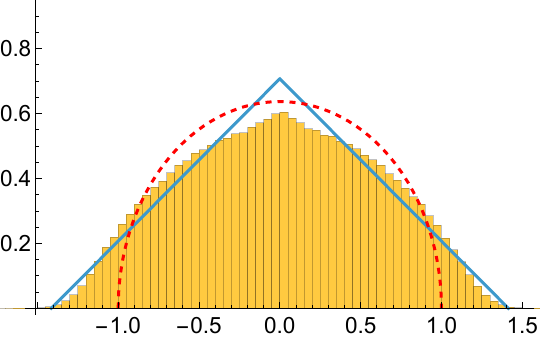} 
	\includegraphics[width=0.3\linewidth]{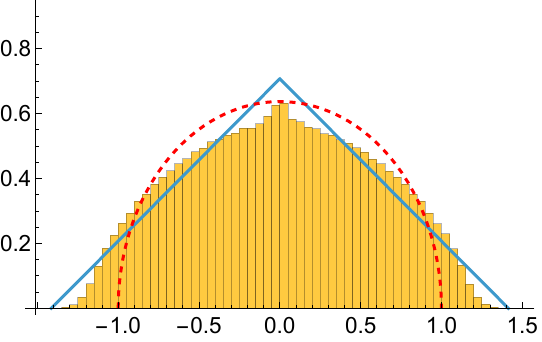}
	\includegraphics[width=0.3\linewidth]{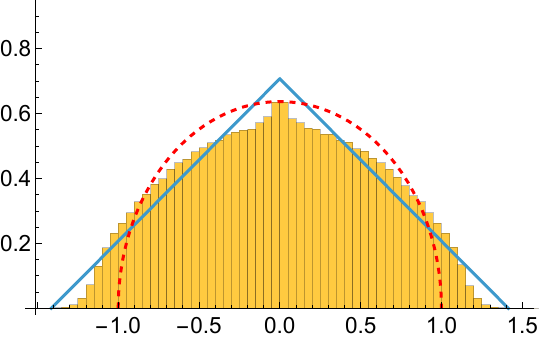}  \\
	\begin{turn}{90}\hspace{20pt}{$\mu=1/2$}\end{turn}
	\includegraphics[width=0.3\linewidth]{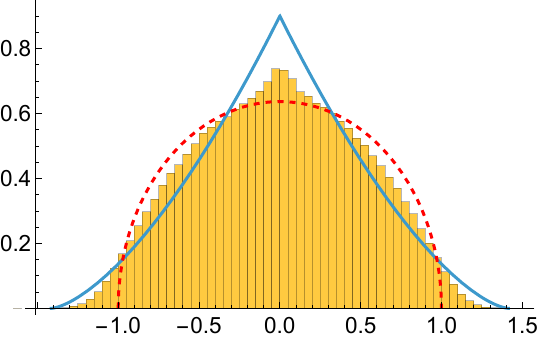} 
	\includegraphics[width=0.3\linewidth]{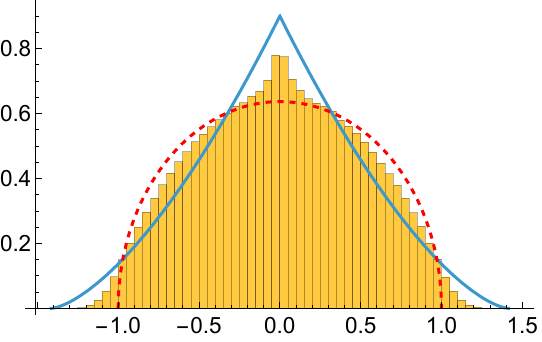}
	\includegraphics[width=0.3\linewidth]{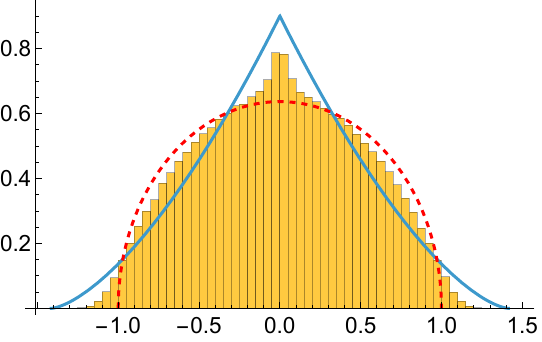}\\  
	\caption{We generated numerically the stationary ensemble of the $\beta$-IRTMP
process in Eq. (\ref{prob.Htridiag}) by drawing independently the diagonal entries from Eq. (\ref{diag_pdf.1})
and off-diagonal entries from Eq. (\ref{off_diag_pdf.1}) using the inverse CDF method described in the text.
We then diagonalized this $N\times N$ random matrix to obtain the histogram of its
$N$ eigenvalues. We plot the
rescaled average eigenvalue density (rescaled by a factor $\sqrt{N}$, respectively $\sqrt{2N}$)  for $N=20$ (the histogram averaged
over $50000$ samples). We chose parameters values 
$\beta=1,\, 2.6,\, 3.1$ (from left to right) and $\mu=1,\, 1/2$ (top, bottom row) respectively. 
We have set $r=D=1$ everywhere.
We compare this eigenvalue density of this $\beta$-IRTMP
process with the known analytical density for the $\beta$-SRTMP process
(shown by blue curves, cf. Eq. (\ref{eq:main-density-special})), 
where we rescaled the density by $1/\sqrt{N}$ in the top row and by $1/\sqrt{2N}$ in the bottom row,
cf. \eqref{eq:main-density}. 
For comparison we give also the semi-circle \eqref{semicircle} (red dashed curve), 
which is the limiting density without resetting. }
	\label{fig-n20 nmtr50000}
\end{figure}

We study the stationary ensemble, defined via Eqs. \eqref{diag_pdf.1} and \eqref{off_diag_pdf.1} numerically. 
For the off-diagonal elements $H_{k,k+1}(t)$ the probability distributions follow from Eqs. 
\eqref{off_diag_element.1}, \eqref{off_diag_pdf.1} with $\textrm{Prob.}\left[H_{k,k+1}=y\right] 
=\textrm{Prob.}\left[Y_k(r)=2y^2\right] 4|y|$. We use the so called \textit{inverse CDF method} to generate the 
random matrix $H$, cf. \cite{Simulation_book}. Let $x$ be a random variable distributed with respect to some 
density $f(z)$. We calculate its cumulative distribution function (CDF) $F(t)=\int_{-\infty}^tdz f(z)$ for 
several values $t$ and list this as pairs $\{F(t),t\}$, which gives data points of the inverse CDF of $x$. Next, 
we interpolate this function linearly and evaluate it at a uniform distributed random variable in 
$\left[0,1\right]$. Here it is important to note that the inverse CDF can diverge around 0 or 1. Thus, a random 
variable in $\left[0.0001,1-0.0001\right]$ is chosen, where we checked that this cut-off is a good choice. 

In the computation of the CDF two integrals arise in each matrix variable, i.e. the $\tau$-integration for 
$\tau\in[0,\infty)$ and the $x$-integration with $x\in (-\infty,t)$ with $t\in\mathbb{R}$ for the diagonal 
elements, respectively the $y$-integration with $y\in[0,t)$ with $t\in \mathbb{R}_+$ for the off-diagonal 
elements, cf. Eqs. \eqref{diag_pdf.1} and \eqref{off_diag_pdf.1}. In order to evaluate these integrals, we have 
to consider cut-off's for the integration domains, where the result is not significantly changed by the respective choice. We consider 
$\tau\in[0,10]$ and $t\in [-2.5,2.5]$ for the diagonal elements, respectively $t\in[0,6]$ for the off-diagonal 
elements. We observe that especially the integrands for the off-diagonal elements become highly oscillatory for 
large $N$, as $q$ in Eq. \eqref{off_diag_para.1} will be large. We set the parameters $D=r=1$ and vary 
$\beta=1,\, 2.6,\, 3.1$ as well as $\mu=1,\, 1/2$.

It is important to note that in this described procedure for the 
simulation of the tridiagonal matrix $H(t\rightarrow \infty)=H$, the resetting is at independent times for each 
matrix elements rather than simultaneous. Hence this average density in the $\beta$-IRTMP ensemble is
expected to be different from the exact result in Eq. (\ref{eq:main-density}) for the $\beta$-SRTMP process.
For comparison, we 
provide the analytical results on 
the limiting density of eigenvalues in the $\beta$-SRTMP ensemble in Fig. \ref{fig-n20 nmtr50000} 
as the blue curves. We plot 
here the rescaled density in the $\beta$-SRTMP ensemble in Eq. (\ref{eq:main-density-special})
for $\mu=1$ and $\mu=1/2$ (with $r=D=1$).
Additionally, we plot the semi-circle as red dashed curve in Fig. \ref{fig-n20 nmtr50000}, 
which is the limiting density without resetting, i.e.
\begin{equation}
	\rho^{\text{sc}}(x)=\frac{2}{\pi}\sqrt{1-x^2}, \text{ with } x\in[-1,1].\label{semicircle}
\end{equation}
We observe in Fig. \ref{fig-n20 nmtr50000} that the stationary states of the models of independent (histograms), 
respectively simultaneous resetting (blue curves) are very different. While the agreement of the histogram data for independent resetting with the blue analytic curve for simultaneous resetting is relatively good in the tail for small $\beta$, it becomes worse for increasing $\beta$ (from left to right). At the origin, we observe the opposite behaviour, i.e. the average density becomes more peaked for large $\beta$. Computing analytically the average density in this $\beta$-IRTMP ensemble remains an interesting
open problem.

\section{Application to a disordered hopping model in one dimension}
\label{Hopping}

Consider a single quantum particle with nearest neightbour hopping on a ring of $N$ sites in the presence
of disorder. The Hermitian
Hamiltonian, in the site basis, reads
\begin{equation}
\mathcal{H}=\sum_{k=1}^{N-1} H_{k,k+1}\left[\, \vert k \rangle \langle k+1\vert +
\vert k+1 \rangle \langle k\vert\, \right] + \sum_{k=1}^N H_{k,k,}\, \vert k \rangle \langle k\vert \, ,
\label{Hamil.1}
\end{equation}
where the diagonal element $H_{k,k}$ represents the local disorder potential at site $k$ and
$H_{k,k+1}\ge 0$ represents the hopping amplitude from site $k$ to site $k+1$. Thus
the disordered environment is encoded in the $(N\times N)$ symmetric tri-diagonal matrix $H_{k,k'}$,
whose off-diagonal elements are non-negative, while the diagonal elements can be either positive or negative.
If the matrix entries $H_{k,k'}$ are random and independent from each other, the Hamiltonian \eqref{Hamil.1}
represents a simple disordered tight-binding version of the celebrated Anderson model for noninteracting fermions on
a lattice.

Given a particular realization of the random matrix $H_{k,k'}$ modelling the disorder, one can
diagonalize the Hamiltonian $\mathcal H$ and obtain its real energy eigenvalues $\vec{x}$.
These eigenvalues will fluctuate from sample to sample of the disorder. Given a realization
of the disorder $\{H_{k,k'}\}$, one of the basic
fundamental quantity is the partition function at a finite temperature $T$ defined as
\begin{equation}
Z\left (T|\{H_{k,k'}\} \right)= \sum_{i=1}^N e^{-x_i/(k_B T)}\, ,
\label{PF.1}
\end{equation}
where $k_B$ is the Boltzmann constant. In the `annealed' version of disorder, the thermodynamic
behavior is captured by averaging directly the partition function over disorder and
compute its logarithm as the annealed free energy
\begin{equation}
F_{\rm annealed}(T)= -k_B\, T\, \ln \left[\, \overline{Z}\,\right]\, ,
\label{ann.1}
\end{equation}
where $\overline{..}$ represents an average over the disorder
\begin{equation}
\overline{Z}= \int dH\, {\rm Prob.}(H)\, Z\left (T|\{H_{k,k'}\} \right)\, .
\label{av.1}
\end{equation}
In contrast, in the `quenched' version of disorder, one first computes the free energy associated
with a fixed realization of disorder and then averages this free energy over the disorder, i.e.,
\begin{equation}
F_{\rm quenched}(T)= -k_B\, T\, \overline{ \ln Z\left (T|\{H_{k,k'}\} \right)}\, .
\label{quenched.1}
\end{equation}
Thus, to compute either the annealed or the quenched average of free energy, one needs to know
the distribution of the partition function $Z\left (T|\{H_{k,k'}\} \right)$
in Eq. (\ref{PF.1}) as the disorder varies. Annealed averaging is typically much simpler compared to
the quenched case. In either case, if one knows
the joint distribution of the eigenvalues $P(\vec x\equiv \{x_1,x_2,\ldots, x_N\})$ of $\mathcal H$, one
can, in principle, obtain the distribution of $Z\left (T|\{H_{k,k'}\} \right)$ since
this is just a linear statistics of eigenvalues as in Eq. (\ref{PF.1}).
More precisely, the distribution of the partition function can be expressed as
\begin{equation}
\tilde{P}(z)= {\rm Prob.}[Z=z] = \int d\vec x\, P(\vec x)\, 
\delta\left(\sum_{i=1}^N e^{-x_i/(k_B\, T)}- z \right)\, .
\label{PF_dist.1}
\end{equation}
If one can compute $\tilde{P}(z)$,  then computing even the quenched average in Eq. (\ref{quenched.1}) is
possible.

One can, of course, sample such a random symmetric tridiagonal matrix $H_{k,k'}$ from any ensemble. For example,
the standard Dumitriu-Edelman $\beta$-ensemble represents a possible choice, where the diagonal
entries are sampled from Gaussians and the off-diagonal elements from $\chi^2$ distribution with
a row-dependent parameter. In fact, one can now introduce a fictitious `time' $t$ parametrizing
the matrix $H_{k,k'}(t)$ and can have a family of disorder ensembles parametrized by $t$.
One example of this is the matrix $H_{k,k'}(t)$ corresponding to the matrix tridiagonal process ($\beta$-TMP)
introduced in this paper, where the diagonal elements perform independent Ornstein-Uhlenbeck processes
as a function of `time' $t$, while the off-diagonal elenents perform independent CIR processes with
a row-dependent parameter. The stationary limit ($t\to \infty$) of this ensemble precisely coincides
with the standard $\beta$-ensemble of Dumitriu-Edelman.

Another choice of disordered ensembles would be to consider the $\beta$-SRTMP process 
where the matrix elements evolving by the $\beta$-TMP process are simultaneously reset to the
origin with rate $r$.
In the limit $t\to \infty$, the matrix $H_{k,k'}(t\to \infty)$ reaches a stationary
state with joint distribution of entries given in Eqs. (86) and (87). In this case,
we know the joint distribution of eigenvalues $P(\vec x)$ explicitly as given in Eq. (91) of the
draft. It would then be nice to compute both the annealed and the quenched free energies, given
respectively in Eqs. (\ref{ann.1}) and (\ref{quenched.1}), for this disordered model.

\vskip 0.2cm

\noindent {\bf Exact annealed free energy.}
In fact, the annealed free energy can be explicitly computed as follows. To compute $F_{\rm annealed}(T)$
in Eq. (\ref{ann.1}), we need to first compute the disorder averaged partition function in
Eq. (\ref{av.1}). Taking average of Eq. (\ref{PF.1}) over disorder, we get
\begin{equation}
\overline {Z\left (T|\{H_{k,k'}\} \right)}= \sum_{i=1}^N \int dx_1\, dx_2\ldots dx_N
e^{-x_i/(k_B T)}\, P(x_1,x_2,\ldots, x_N)\, .
\label{ann.2}
\end{equation}
We now use the fact that the average density of eigenvalues can be expressed as the one-point marginal of the
joint distribution of eigenvalues, i.e.,
\begin{equation}
\rho_N(x) = \frac{1}{N}\, \left\langle \sum_{i=1}^N \delta(x-x_i) \right\rangle=
\int dx_2\,dx_2\ldots dx_N\, P(x, x_2, x_3,\ldots x_N) 
\label{ann.3}
\end{equation}
where $\langle \cdot \rangle$ denotes the averaging over the joint distribution and we used the exchangeability
of the joint PDF. Using (\ref{ann.3}) in Eq. (\ref{ann.2}) gives
\begin{equation}
\overline {Z\left (T|\{H_{k,k'}\} \right)}= N\, \int dx\, \rho_N(x)\, e^{-x/(k_B\, T)}\, .
\label{ann.4}
\end{equation}
Hence all we need to know is the average density $\rho_N(x)$ which has already been in computed
in Eqs. (31) and (32) of Ref.~\cite{BMS2025} in the large $N$ limit, for all $\beta$ and $r$. It is given in Eq. \eqref{eq:main-density}.

Substituting Eq. (\ref{eq:main-density}) in (\ref{ann.4}) gives
\begin{equation}
\overline {Z\left (T|\{H_{k,k'}\} \right)}= N\, \int_{-\sqrt{2}}^{\sqrt{2}} dy\, e^{-\alpha\, y}\,
f(y,\nu)\, , \quad\quad {\rm where}\quad 
\alpha= \frac{1}{k_B\, T}\, \sqrt{\frac{N\, D}{\mu}}\, , \quad {\rm and}\quad \nu=\frac{\mu}{r}\, ,
\label{ann.5}
\end{equation}
with the scaling function $f(y,\nu)$ given in \eqref{eq:main-f}. In general, it is not easy to
perform this integral explicitly for arbitrary $\nu$. However, for $\nu=1$, using
Eq. (\ref{eq:main-density-special}), we get $f(y,1)= 1/\sqrt{2}- |y|/2$ for $y\in [-\sqrt{2},\sqrt{2}]$.
In this case, performing the integral in \eqref{ann.5} we get
\begin{equation}
\overline {Z\left (T|\{H_{k,k'}\} \right)}\Big|_{\gamma=1}= N\,
\left[ \frac{\sinh(B)}{B}\right]^2\, , 
\label{ann.6}
\end{equation}
where
\begin{equation}\label{defB}
	B= \frac{\alpha}{\sqrt{2}}= \frac{1}{k_B T}\,
	\sqrt{\frac{N\, D}{2\, \mu}}\, .
\end{equation}
Consequently, the annealed free energy, from Eq. (\ref{ann.1}) is given explicitly for $\nu=1$ as
\begin{equation}
F_{\rm annealed}(T)\Big|_{\gamma=1}= -k_B T\, \left[ \ln N
+ 2\, \ln \left(\frac{\sinh(B)}{B}\right)\right]\, , 
\label{ann.7}
\end{equation}
with $B$ given in Eq. \eqref{defB}.
In the limit of large $N$, the parameter $B\sim \sqrt{N}$ becomes large and hence the annealed free
energy scales as
\begin{equation}
F_{\rm annealed}(T)\Big|_{\gamma=1}\approx -\sqrt{\frac{2 N\, D}{\mu}}\, .
\label{ann.8}
\end{equation}
Thus, interestingly, to leading order in the large $N$ limit, the free energy becomes independent of temperature $T$
and scales subextensively as $\sqrt{N}$.
In fact, the rhs of \eqref{ann.8} is just the 
minimum eigenvalue $x_{\rm min}=-\sqrt{2ND/\mu}$ in the large $N$ limit.
This is consistent with the fact that taking the large $N$ limit, 
is equivalent to taking the $T\to 0$ limit, since
the relevant parameter is $B= \sqrt{ND/(2\mu)}/(k_B T)$. At $T=0$, we expect the ground state, i.e.,
the minimum eigenvalue to dominate the free energy. Thus, in the context of disordered system, this
means that the free energy, in the large $N$ limit, is dominated by disorder (or energy) only and the
temperature/entropy plays no role.

Another solvable case is when $r=0$, or equivalently as $\nu\to \infty$. This corresponds to
modelling the disorder by the the classical Dumitriu-Edelman $\beta$-ensemble.
In this case, it follows from
the $\nu\to \infty$ limit of Eq. (\ref{eq:main-f}) that the average density 
converges to the Wigner semi-circular form
\begin{equation}
f(z,\nu\to \infty)= \sqrt{\frac{2}{\pi^2}}\, \left[1- \frac{z^2}{2}\right]^{1/2}\, .
\label{wigner.1}
\end{equation}
Substituting this result in Eq. (\ref{ann.5}), one gets the average partition function
\begin{equation}
\overline {Z\left (T|\{H_{k,k'}\} \right)}\Big|_{\nu\to \infty}= \frac{N}{B}\,
I_1(2\, B)\, ,  
\label{ann_semicircle.1}
\end{equation}
with $B$ given in Eq. \eqref{defB} and $I_\nu(z)$ is the modified Bessel function of the first kind with index $\nu$. Consequently,
the annelaed free energy can also be computed explicitly from Eq. (\ref{ann.1}) and one sees
qualitatively similar behavior as $\gamma=1$. Computing 
the quenched free energy in \eqref{quenched.1}) for this
disordered ensemble remains an interesting open problem.

\section{Conclusion}
\label{Conclusion}

In this paper we introduced a symmetric tridiagonal matrix-valued process ($\beta$-TMP) $H(t)$ whose diagonal 
entries $H_{k,k}(t)$ evolve independently via an Ornstein-Uhlenbeck process starting at the origin and the 
off-diagonal entries $H_{k,k+1}(t)$ evolve independently via the Cox-Ingersoll-Ross (CIR) process, starting at 
the origin and with parameters that depend on the row index $k$. The joint distribution of the entries of the 
matrix can be computed exactly at all times and moreover, the joint distribution of its $N$ real eigenvalues can 
also be computed exactly at all times and is given by the $\beta$-ensemble of Dumitriu-Edelman up to a simple 
time-dependent rescaling factor with arbitrary Dyson index $\beta>0$. We then subject this time-evolving matrix 
valued process to stochastic resetting with rate $r$ in two different settings: (i) when the matrix entries are 
simultaneously reset to the origin with rate $r$ ($\beta$-SRTMP process) and (ii) when the matrix entries are 
independently reset to the origin with rate r ($\beta$-IRTMP process). In the $\beta$-SRTMP process, the matrix 
elements get strongly correlated in the stationary state, while in the $\beta$-IRTMP process the matrix elements 
remain independent at all times, including in the stationary state. We showed that the joint distribution of the 
eigenvalues of the $\beta$-SRTMP process at long times coincides with the joint distribution of the positions of 
the resetting Dyson Brownian motion for arbitrary $\beta>0$. While such an underlying resetting matrix-valued 
process was known previously~\cite{BMS2025} only for three special values of the Dyson index $\beta=1,2$ and 
$4$, our construction of the $\beta$-SRTMP process here generalises the resetting matrix-valued process to 
arbitrary $\beta>0$.
 
We have shown that for the $\beta$-SRTMP stationary ensemble, the joint distribution of eigenvalues can be 
computed analytically for all $\beta>0$. In contrast, for the $\beta$-IRTMP stationary ensemble, while the 
matrix ensemble can be generated numerically very easily (since the matrix entries are independent), computing 
the joint distribution of eigenvalues is not easy and remains an open problem. Even computing the average 
density of eigenvalues analytically in the stationary $\beta$-ITRMP process seems hard. In fact, we computed 
numerically the average density of eigenvalues in the $\beta$-IRTMP stationary ensemble and compared them to the 
analytical density for the $\beta$-SRTMP process and demonstrated that they are quite different from each other.

We also provided a simple and concrete application of the $\beta$-SRTMP ensemble to study the partition function 
of a tight-binding quantum Hamiltonian representing a single particle hopping on a one-dimensional lattice at 
finite temperature with disordered hopping rates drawn from the $\beta$-SRTMP stationary ensemble. We computed 
exactly the annealed free energy of this tight-binding model. Computing analytically the quenched free energy 
also remains an interesting open problem for future. In addition, if the disordered hopping rates are drawn from 
the stationary $\beta$-IRTMP ensemble, computing the free energy, both annealed and quenched, remains a 
challenging open problem.

\subsection*{Acknowledgements}

The work of Gernot Akemann and Patricia P\"a{\ss}ler were partly funded by the 
Deutsche Forschungsgemeinschaft (DFG) grant SFB 1283/2 2021 – 317210226.
The work of Satya N. Majumdar was partially funded by ANR Grant No. ANR-
23-CE30-0020-01 EDIPS.
We thank the organisers of the workshop "Random Matrix Theory for Ecology, Economics, Finance and Statistical Physics" of the research programme 
"The Lush World of Random Matrices" at the Centre for Interdisciplinary Research ZiF (Bielefeld) for a 
stimulating atmosphere that initiated this work.\\

\end{document}